\documentclass[prb,10pt,twocolumn,superscriptaddress,footnoteinbib]{revtex4}
\usepackage{amsmath}
\usepackage{xcolor}
\definecolor{deepgreen}{RGB}{0,100,0}
\usepackage{latexsym}
\usepackage{amssymb}
\usepackage{graphics,epstopdf}
\usepackage[colorlinks=true, citecolor=blue, urlcolor=blue ]{hyperref}
\usepackage{epsf,graphics,graphicx}
\usepackage{soul}

\begin{document}

\title{Beyond-ballistic transport in an open quantum ring}

\author{Moumita Patra}

\email{moumita.patra19@gmail.com}

\affiliation{School of Physical Sciences, Indian Association for the Cultivation of Science, 2A \& 2B Raja S. C. Mullick Road, 
Kolkata-700 032, India}

\affiliation{Department of Physics, Indian Institute of Science Education and Research, Pune-411 008, India}

\author{Bijay Kumar Agarwalla}

\email{bijay@iiserpune.ac.in}

\affiliation{Department of Physics, Indian Institute of Science Education and Research, Pune-411 008, India}

\author{Santanu K. Maiti}

\email{santanu.maiti@isical.ac.in}

\affiliation{Physics and Applied Mathematics Unit, Indian Statistical Institute, 203 Barrackpore Trunk Road, Kolkata-700 108, India}

\begin{abstract}

In an open quantum ring (OQR), an asymmetric ring-to-electrode configuration, where
the upper and lower arms have unequal lengths, generates antiresonances in the junction
transmission spectrum around the doubly degenerate eigenenergies of the isolated ring
Hamiltonian. The asymmetric OQR also gives rise to a net circular current transmission within the ring,
specifically around these doubly degenerate eigenenergies. We investigate the system-size
scaling properties of the transmission within the ring and the overall junction transmission
of an OQR. Ballistic transport refers to the unhindered flow of charge carriers within a conductor,
where transmission is independent of the system size.
Here, we find {\it beyond-ballistic} behavior, characterized by an anomalous increase of the
transmission with increasing system size, near both the degenerate and non-degenerate
eigenenergies of the ring Hamiltonian, depending on the ring-to-electrode configuration.
This phenomenon is unique to OQRs and is associated with the quantum interference effect
between two counter-propagating electronic waves with nearly equal and opposite momenta.
Consequently, there is no equivalent phenomenon in open quantum junctions with linear
conductors.

\end{abstract}

\maketitle

\section{Introduction}

The Drude model, proposed about 100 years ago, describes electronic transport in metals~\cite{drude}.
Since then, the nature of electron transport has remained
a topic of primary interest across a broad area of research, and yet fresh mysteries and revelations
continue to arise. Three scenarios are typically observed in the context of charge-carrier
transport in low-dimensional systems~\cite{trans1,trans2,trans3}. Ballistic transport
occurs in a perfect conductor, where the conductance
$G$ is independent of $N$, and charge carriers flow without any restriction. In an insulator,
$G$ decreases exponentially with system size $N$, i.e., $G \sim e^{-\lambda N}$ with
$\lambda > 0$, where $\lambda$ is the inverse localization length. On the other hand, diffusive
transport appears in the presence of disorder, where $G$ scales as $N^{-1}$.

The nature of electron transport is studied primarily in isolated systems~\cite{bd1,bd2,bd3,bd4}.
For example, the well-known Anderson localization~\cite{al1,al2}, a problem of long-standing
interest in condensed matter physics~\cite{al3}. In a tight-binding lattice,
the system wave function changes from being extended (metal) to exponentially localized
(insulator) in the presence of an infinitesimal random disorder. On the other hand, in the
Aubry-Andr\'{e}-Harper (AAH) model, the metal-to-insulator transition arises from the presence of
an incommensurate potential of finite strength~\cite{AAH1,AAH2}.

In the context of transport, another important variant is the open quantum system where the
system of interest is coupled to multiple environments that drive current through it. Often,
in such cases, the system-size dependences of current, conductance, and transmission probability
are studied for linear geometries~\cite{ot1,ot2,ms}.
\begin{figure}[ht]
{\centering\resizebox*{8cm}{3.75cm}{\includegraphics{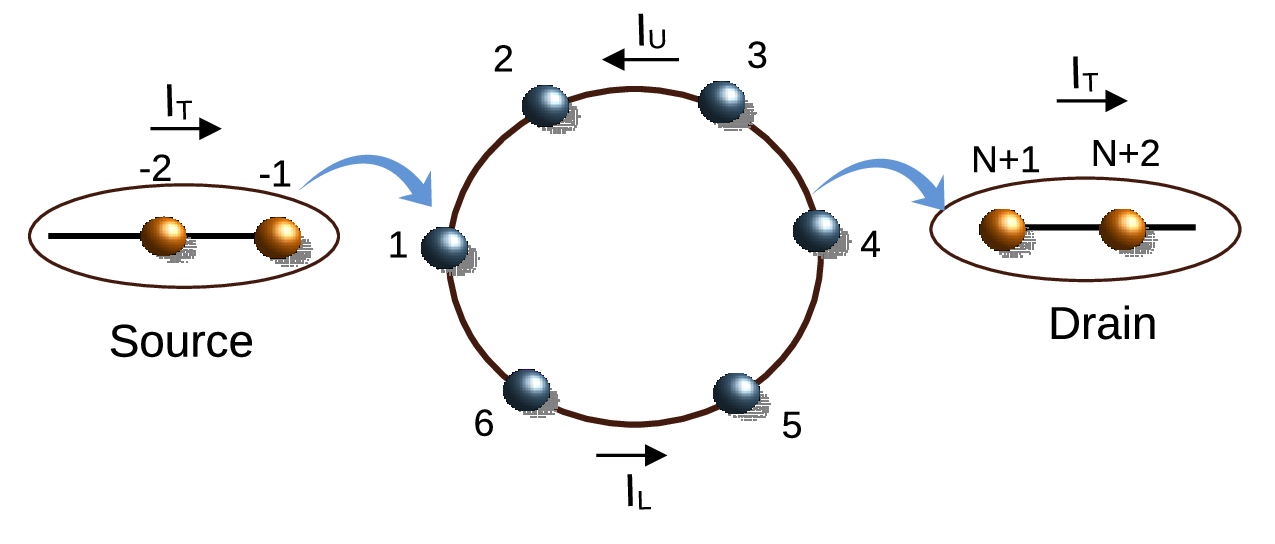}}\par}
\caption{(Color online). Schematic view of a tight-binding quantum ring attached to two semi-infinite electrodes. $I_U$ and $I_L$
represent the currents in the upper and lower arms of the ring, respectively. $I_T$ refers to the overall junction current.}
\label{fig1}
\end{figure}
To the best of our knowledge, no studies have examined the $N$-dependence of transport in open quantum rings (OQRs). 
In this paper, {\em we investigate the system-size scaling of electron transport in a tight-binding fermionic open quantum ring,
a problem that has not been addressed so far in the literature}.

An OQR exhibits unique transport properties compared to linear conductors. In an OQR,
the current inside the conductor and the overall junction current can show markedly different
quantitative and qualitative behavior~\cite{cir1,cir2,cir3,cir4,Nitzan1,cir6,SKM,cir8,cir9,cir10}, 
in contrast to a linear conductor where, due to charge conservation, the current inside the conductor
must equal the current flowing through the external circuit. Consider an OQR as shown in Fig.~\ref{fig1},
consisting of six atomic sites and symmetrically connected to the source (S) and drain (D) electrodes,
so that the lengths of the upper and lower arms are equal. The zero-temperature bond current within the ring and the
overall junction current can be written as~\cite{cir6,sd1,sd2,mpprb}
$I_{i \rightarrow i+1}(V) = \frac{2 e}{h} \int T_{i \rightarrow i+1}(E)\, dE$ and $I_T(V) = \frac{2 e}{h} \int T(E)\, dE$,
respectively, where $e$ is the electron charge and $h$ is Planck’s constant.
Here, $T_{i \rightarrow i+1}(E)$ and $T(E)$ denote the bond transmission coefficient and
transmission probability of junction current, respectively.
{\it The junction transmission probability $T(E)$ is always positive and bounded above by unity in the present single-channel model,
whereas $T_{i \rightarrow i+1}(E)$ can take either positive or negative values and is unbounded in an OQR.}

Because of current conservation, the bond currents along any arm are identical. Thus, for the upper arm of
Fig.~\ref{fig1}, $I_{1 \rightarrow 2} = I_{2 \rightarrow 3} = I_{3 \rightarrow 4} \equiv I_U$, and similarly for the
lower arm, $I_{4 \rightarrow 5} = I_{5 \rightarrow 6} = I_{6 \rightarrow 1} \equiv I_L$. The overall junction current
must satisfy $I_T = I_L - I_U$. In our sign convention, currents flowing anticlockwise are taken to be positive.
Under non-equilibrium conditions, $I_U$ and $I_L$ flow from the electrode with higher chemical potential to the one
with lower chemical potential. Thus, for a symmetric ring, $I_U$ is negative, $I_L$ is positive, and
$I_U = -I_L$. However, for an asymmetric ring, both currents can flow in the same direction, giving rise to a
net bias-driven circulating current which is the average of the branch currents~\cite{Nitzan1,Nitzan,zero1,zero2}.

In the present work, we investigate the system-size scaling of the bond transmission coefficients,
the net circular transmission coefficient, and the overall junction transmission probability in an
open quantum ring (OQR). For the symmetric ring-to-electrode configuration, the net circular
transmission coefficient vanishes, and therefore we present only the bond transmission coefficients
and the junction transmission probability. For the asymmetric configuration, we present all three
quantities. We find that the bond transmission coefficients, the net circular transmission
coefficient, and the junction transmission probability exhibit the same qualitative dependence on
the system size. Moreover, in the asymmetric configuration, the bond and net circular transmission
coefficients show nearly identical qualitative and quantitative behavior around the degenerate
eigenenergies of the bare ring.

Unlike a linear chain, a ring Hamiltonian possesses doubly degenerate energy eigenvalues, arising from the
periodic boundary condition of the ring, since eigenstates with momenta $+k$ and $-k$ share the same energy.
When the ring is connected to electrodes in a geometry where the upper and lower arms have unequal lengths,
the {\it transmission resonance associated with these doubly degenerate eigenenergies splits into two resonant
peaks}. In such an asymmetric ring, the overall junction transmission exhibits two sharp
resonance peaks separated by a zero (anti-resonance) in the vicinity of the doubly degenerate
eigenenergies of the isolated ring. This behavior originates
from quantum-mechanical interference, and the corresponding poles and zeros in the conductance have been
well studied~\cite{zero1,zero2}. The circular current appears in the vicinity of these doubly degenerate
eigenenergies.

We find that, although the transport current and the circular current respond differently to variations in
bias voltage, disorder, and other external parameters, they display the same qualitative dependence on system
size. Similar to linear conductors, transport is ballistic except near the degenerate energy levels associated
with any quantum ring (QR). For a symmetric ring, the transmission increases with system size $N$ at the band edge.
{\it The growth of transmission with system size that exceeds the ballistic expectation is referred to here as
beyond-ballistic transport.}

We also observe beyond-ballistic transport around the anti-resonances of an asymmetric ring. Furthermore,
the extent of the beyond-ballistic regime can be tuned by adjusting the ring--electrode coupling strength.
This anomalous scaling arises from the general property that quantum rings whose system sizes are related
by an integer multiple share the same eigenenergies, leading to transmission peaks of identical height
while the peak width decreases with increasing system size. This repetition of eigenstates is an intrinsic
property of quantum rings and has no analogue in linear junctions. Consequently, such anomalous scaling does
not occur in linear conductors.

In realistic situations, the transport characteristics of mesoscopic systems are influenced by various physical factors
such as structural disorder, finite temperature, and dephasing processes. These effects generally tend to suppress
quantum interference and modify the scaling behavior. While the present study focuses on coherent transport at zero
temperature for clarity, the role of dephasing has been examined following the approach described in
Ref.~\cite{mpprb}. A brief discussion on the influence of finite temperature and decoherence has also been
included. Moreover, in higher-dimensional systems, the presence of subband structures and inter-subband coupling may
alter the interference pattern and scaling response. For comparison, we have analyzed a two-dimensional square
lattice configuration, where conventional ballistic transport behavior is recovered, confirming that the anomalous
scaling reported here is a genuine feature of the strictly one-dimensional open quantum ring geometry. In addition,
to connect these theoretical predictions with realistic conditions, we include a discussion on the experimental
perspective for observing circular currents and the associated beyond-ballistic transport in open quantum rings.

The remainder of the paper is organized as follows. In Sec.~II, we outline the wave-guide formalism used to compute
the transmission coefficient and probability. Section~III presents the key numerical results and their analysis. The possible
experimental perspective is discussed in Sec.~IV. Finally, Sec.~V summarizes our main findings.

\section{Theoretical Framework}

The tight-binding Hamiltonian $H$ for the OQR (see Fig.~\ref{fig1}) can be written as the sum of sub-Hamiltonians as
\begin{equation}
H = H_{\mbox{\tiny Ring}} +
H_{\mbox{\tiny Source}} + H_{\mbox{\tiny Drain}} + H_{\mbox{\tiny Tunneling}}.
\label{hamil}
\end{equation}
\noindent
The forms of these sub-Hamiltonians are
\begin{equation}
H_{\mbox{\tiny Ring}} = \sum\limits_{i=1}^{N} \epsilon_i c_i^{\dagger} c_i + \sum\limits_{i=1}^{N} t
\left(c_{i+1}^{\dagger} c_i + c_i^{\dagger} c_{i+1} \right),
\label{Eq1}
\end{equation}
\begin{equation}
H_{\mbox{\tiny Source}} = \sum\limits_{n \le -1} \epsilon_0 a_n^{\dagger} a_n +
\sum\limits_{n \le -1} t_0 \left(a_n^{\dagger} a_{n-1} +
a_{n-1}^{\dagger} a_n \right),
\label{Eq2}
\end{equation}
\begin{equation}
H_{\mbox{\tiny Drain}} = \sum\limits_{n \ge 1} \epsilon_0 b_n^{\dagger} b_n +
\sum\limits_{n \ge 1} t_0 \left(b_n^{\dagger} b_{n+1} +
b_{n+1}^{\dagger} b_n \right),
\label{Eq3}
\end{equation}
\begin{eqnarray}
H_{\mbox{\tiny Tunneling}} & = & t_S \left(c_{N_S}^{\dagger}a_{-1} +
a_{-1}^{\dagger} c_{N_S} \right)\nonumber \\
& & + t_D \left(c_{N_D}^{\dagger}b_1 + b_1^{\dagger} c_{N_D} \right).
\label{Eq4}
\end{eqnarray}
Here, the meanings of the different symbols are as follows.
$\epsilon_0$ and $\epsilon_i$ ($i = 1, 2, \dots, N$) are the on-site potentials of the electrodes and
the ring, respectively. $t_0$ and $t$ are the nearest-neighbor hopping strengths of these components.
The creation operators $a_n^{\dagger}$, $b_n^{\dagger}$, and $c_i^{\dagger}$
are associated with the source, drain, and ring, respectively. $t_S$ denotes the coupling strength between the
ring and the source, while $t_D$ denotes the coupling between the ring and the drain. The site of the ring coupled to the source is labeled
as $N_S$, and the site connected to the drain is labeled as $N_D$. For instance, in Fig.~\ref{fig1}, $N_S = 1$ and $N_D = 4$.

To evaluate the bond and overall junction transmission, we need to calculate the bond current density $J_{i \rightarrow i+1}(E)$ and
the transmission probability $T(E)$. To calculate $J_{i \rightarrow i+1}(E)$ and $T(E)$,
we adopt the wave-guide prescription~\cite{wg1,wg2}, where we solve a set of coupled linear equations
generated from the Schr\"{o}dinger equation $H|\psi\rangle = E|\psi\rangle$, with
\begin{equation}
|\psi\rangle = \left[\sum\limits_{n \le -1} A_n a_n^{\dagger} +
\sum\limits_{n \ge 1} B_n b_n^{\dagger} +
\sum\limits_{i=1} C_i c_i^{\dagger}\right]|0\rangle.
\label{Eq5}
\end{equation}
The coefficients $A_n$, $B_n$, and $C_i$ represent the electronic wave amplitudes
at the $n$-th site of the source and drain electrodes and at the $i$-th site
of the ring, respectively. In terms of the reflection and transmission coefficients, $r$
and $\tau$, the amplitudes $A_n$ and $B_n$ can be written as
$A_n = e^{ik(n+1)a} + r e^{-ik(n+1)a}$ and $B_n = \tau e^{ikna}$, where $k$ is the wave vector and $a$ is the lattice spacing.
We assume a plane-wave incidence of unit amplitude from the source electrode.

The coupled equations are~\cite{wg1,wg2}
\begin{eqnarray}
\left(E - \epsilon_{0}\right)
A_n &=& t_{0}\left(A_{n+1} + A_{n-1}\right), \quad n \le -2, \nonumber \\
\left(E - \epsilon_{0}\right)
A_{-1} &=& t_{0} A_{-2} + t_{S} C_{N_S}, \nonumber \\
\left(E - \epsilon_{0}\right)
B_n &=& t_{0}\left(B_{n+1} + B_{n-1}\right), \quad n \ge 2, \nonumber \\
\left(E - \epsilon_{0}\right) B_1
&=& t_0 B_2 + t_{D} C_{N_D}, \nonumber \\
\left(E - \epsilon_i\right)
C_i &=& t\left(C_{i+1} + C_{i-1}\right)
+ t_S \delta_{i,N_S} A_{-1} \nonumber \\
&& + t_D \delta_{i,N_D} B_1, \quad 1 \le i \le N.
\label{eqn7}
\end{eqnarray}
Solving Eqs.~\ref{eqn7}, we obtain the bond current density and transmission probability using
\[
J_{i \rightarrow i+1}(E) = \frac{2t}{\hbar}\,\mathrm{Im}\!\left(C_i^* C_{i+1}\right)
\]
and
\[
T(E) = |\tau|^2 = |B_1|^2,
\]
respectively.

The transport (junction) current is calculated in terms of transmission probability
using the Landauer formula,
\begin{eqnarray}
I_T(V) & = & \frac{2e}{h}\int_{-\infty}^{+\infty} T(E)\,[f_S(E)-f_D(E)]\, dE,
\label{ref2}
\end{eqnarray}
where $T(E)$ is the two-terminal transmission probability~\cite{cir6,sd1,sd2,SDatta}.
\begin{figure*}
{\centering\resizebox*{13cm}{6cm}{\includegraphics{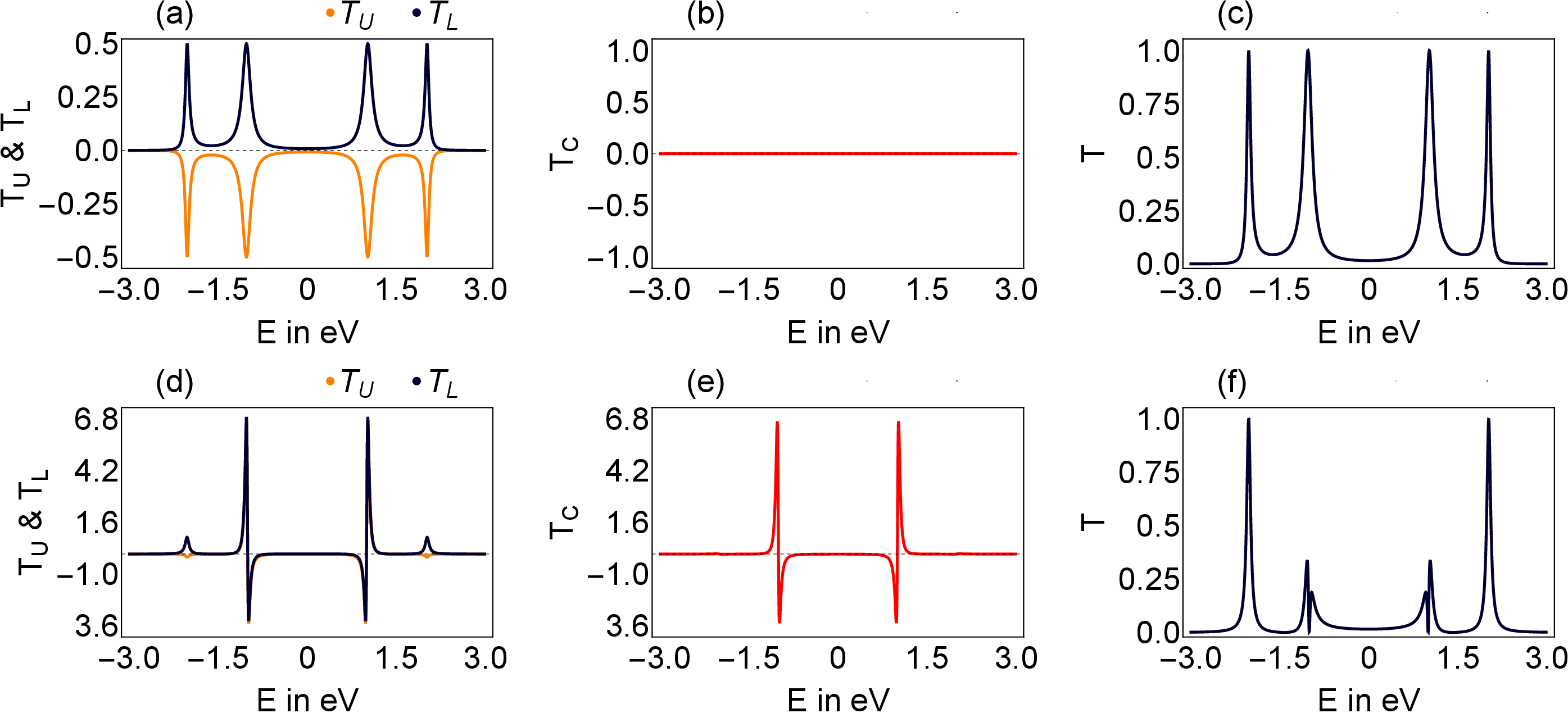}}\par}
\caption{(Color online). Results for a six-site open quantum ring.
The first row corresponds to the symmetric ring-to-electrode configuration,
and the second row to the most asymmetric configuration.
Panels (a) and (d) show the bond transmission coefficients,
$T_U$ and $T_L$, as functions of energy.
Panels (b) and (e) show the corresponding net circular transmission
coefficient, $T_C$, as a function of energy.
Panels (c) and (f) show the junction transmission probability,
$T$, as a function of energy.
The horizontal dashed line marks the zero value of the plotted quantity.}
\label{fig2}
\end{figure*}
The functions $f_S(E)$ and $f_D(E)$ denote the Fermi--Dirac
distribution functions of the source and drain electrodes, respectively.

Analogous to the two-terminal transmission probability, Nitzan and
co-workers~\cite{Nitzan1, cir6, Nitzan, RaiGalperin2012} introduced the bond transmission coefficient to describe
the current flowing through an individual bond of the ring,
\begin{eqnarray}
T_{i\rightarrow i+1}(E)& = &\frac{J_{i\rightarrow i+1}(E)}{J_{\mathrm{incoming}}},
\label{ref3}
\end{eqnarray}
where $J_{\mathrm{incoming}}$ is the incident current density. Unlike the conventional
transmission probability, the bond transmission coefficient
$T_{i\rightarrow i+1}(E)$ can exceed unity and may also take negative
values~\cite{Nitzan1, cir6, Nitzan, RaiGalperin2012, Marathe2010, Nakanishi2001}.

The net circular transmission coefficient is defined as the average of all bond
transmission coefficients,
\begin{eqnarray}
T_C(E) & = & \frac{1}{N}\sum_{i=1}^{N}T_{i\rightarrow i+1}(E).
\label{ref3a}
\end{eqnarray}
Like the bond transmission coefficient, $T_{C}(E)$ can also exceed unity
and may take negative values.

The corresponding circular current is then obtained from the Landauer formula as
\begin{eqnarray}
I_C(V)& = & \frac{2e}{h}\int_{-\infty}^{+\infty}T_C(E)\,[f_S(E)-f_D(E)]\, dE.
\label{ref4}
\end{eqnarray}
\noindent
The transport and bond currents satisfy the current-conservation relation,
\begin{equation}
I_T(V)=I_L(V)-I_U(V).
\label{ref1}
\end{equation}

\section{Numerical RESULTS AND DISCUSSION}

Unless otherwise specified, all on-site energies are set to zero throughout the computations. 
The hopping parameter inside the ring is set to $1\,$eV. 
The source is always connected to the first site of the ring, i.e., $N_S=1$, and the drain position $N_D$ can vary,
as mentioned in the appropriate parts of the discussion. 
The inter-atomic spacing ($a$) is considered to be $1\,$\AA~\cite{mpS}. 
The source-to-ring coupling $t_S$ and ring-to-drain coupling $t_D$ are always assumed to be identical; 
for simplification, we denote them by a single parameter $t_C$. 

\subsection{Energy dependence of the transmissions}

Before studying the system size scaling of the transmission probability (or coefficient), let us first concentrate on
the transmission–energy characteristics. In Fig.~\ref{fig2}, we plot the bond and circular transmission coefficients
and overall junction transmission as functions of energy. The hopping parameter
for the electrodes is $2\,$eV, and the ring-to-electrode coupling is $0.5\,$eV.
$T_U$ and $T_L$ are the transmission coefficients for the upper and lower arms of the ring, respectively.
The energies at which the peaks and dips appear in the transmission spectrum are associated with
the energy eigenvalues of the isolated ring Hamiltonian (Eq.~\ref{Eq1}). The energy dispersion relation for
the isolated ring is $E = 2t \cos(ka)$, where $k = 2\pi m/(Na)$. The integer $m$
runs between $-N/2 \le m < N/2$. Therefore, for $k$ and $-k$, the system
has the same energy. The non-degenerate energy levels correspond to $m = 0$ for odd $N$
and $m = -N/2,\,0$ for even $N$. For example, in Fig.~\ref{fig2} we choose $N = 6$, which implies
$m = -3, -2, -1, 0, 1, 2$. Therefore, degeneracies appear at $m = \pm 2$ and $\pm 1$, whereas the
energy levels corresponding to $m = -3$ (i.e., $E = 2\,$eV) and $m = 0$ (i.e., $E = -2\,$eV) remain
non-degenerate. Thus, the band edges are non-degenerate and the other eigenenergies
of a quantum ring with even $N$ are doubly degenerate.

In the symmetric ring-to-electrode configuration, shown in the first row of Fig.~\ref{fig2}(a),
four resonant peaks are observed in the bond transmission coefficients, corresponding to the
eigenenergies of the isolated (bare) ring. Since the ring-to-electrode coupling preserves the symmetry of the ring,
the transmission resonance associated with each doubly degenerate eigenenergy of the isolated ring remains unsplit.
Consequently, the resonances associated with the degenerate eigenenergies of the bare ring remain
unsplit. The bond transmission coefficients satisfy $T_U=-T_L$, as shown in
Fig.~\ref{fig2}(a). Consequently, the net circular transmission coefficient, defined as the
average of all bond transmission coefficients, vanishes throughout the energy spectrum
[Fig.~\ref{fig2}(b)]~\cite{zero1,zero2,Nitzan1,Nitzan}. The junction transmission probability is shown in Fig.~\ref{fig2}(c).
In this case, $T=T_L-T_U=2|T_U|=2|T_L|$.

A qualitatively different behavior is observed for the most asymmetric ring-to-electrode
configuration, as shown in Figs.~\ref{fig2}(d)--(f). Here, six resonant peaks appear in both the
bond transmission coefficients and the junction transmission probability. It is important to
emphasize that the eigenenergies of the isolated (bare) ring remain unchanged. Instead, the
asymmetric ring-to-electrode coupling modifies the transmission spectrum of the open ring. Around
each doubly degenerate eigenenergy of the bare ring (for example, $E=\pm1\,$eV), the transmission
spectrum exhibits two resonant peaks separated by an anti-resonance. For the present asymmetric
ring-to-electrode configuration, the anti-resonance occurs precisely at the corresponding doubly
degenerate eigenenergy of the isolated ring. We also numerically verify that the transmission
coefficients satisfy $T=T_L-T_U$, consistent with current conservation.

Unlike the symmetric case, the bond transmission coefficients $T_U$ and $T_L$ have the same sign
in the vicinity of the doubly degenerate eigenenergies of the bare ring. Consequently, the net
circular transmission coefficient becomes finite, as shown in Fig.~\ref{fig2}(e). In contrast,
around the non-degenerate band-edge eigenenergies of the bare ring, the net circular transmission
coefficient remains negligibly small. Therefore, in the vicinity of the doubly degenerate
eigenenergies, the net circular transmission coefficient exhibits nearly the same qualitative
and quantitative behavior as the bond transmission coefficients.

The bond and circular transmission coefficients can take both positive and negative values.
These sign changes do not correspond to negative two-terminal conductance or negative
resistance. Instead, they simply reflect the direction of the local current within the
ring. In an open quantum ring, current splitting at the junctions forces one arm to
\begin{figure}[ht]
{\centering \resizebox*{8cm}{6cm}{\includegraphics{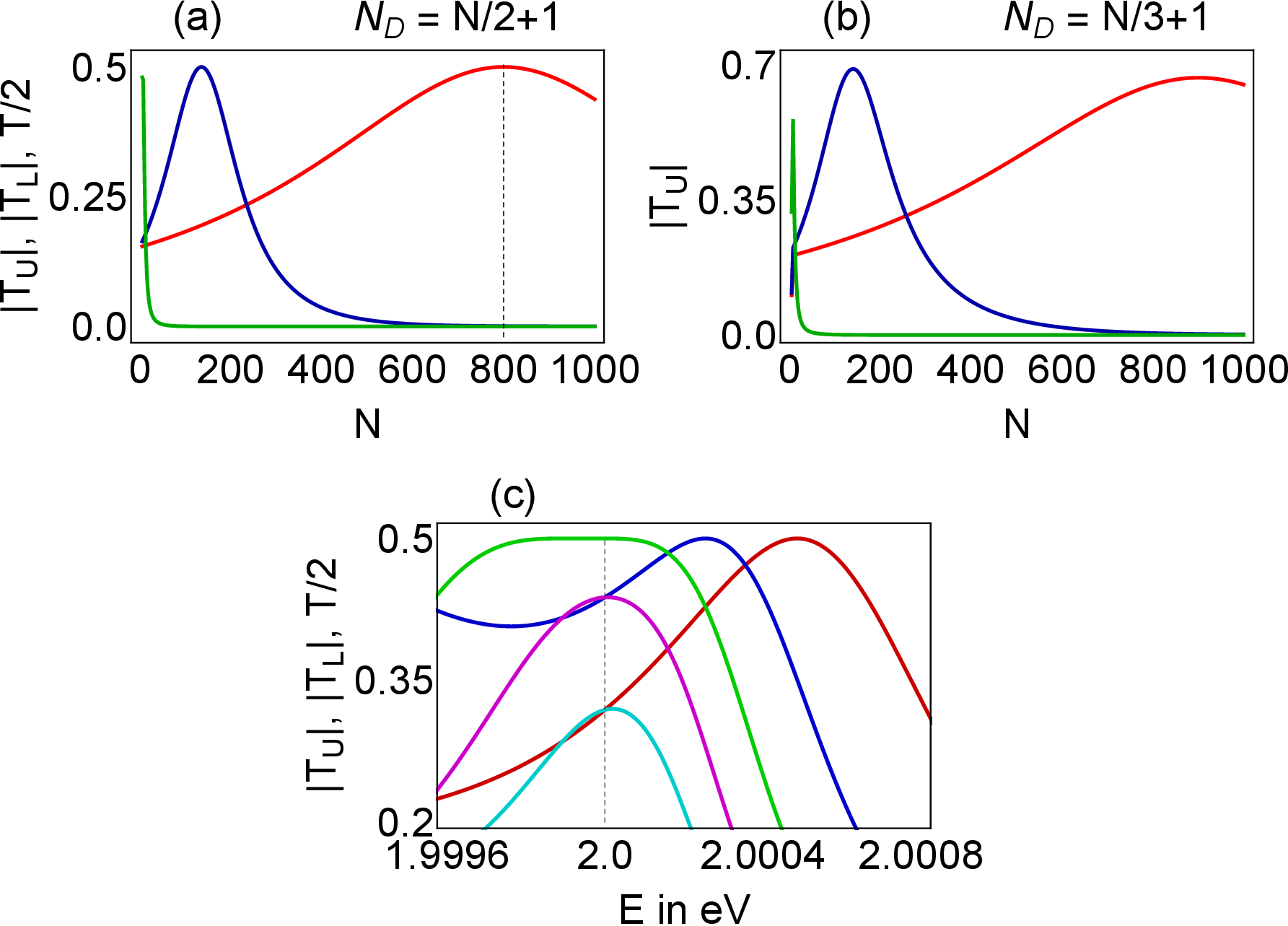}}\par}
\caption{(Color online). In (a) and (b), $|T_U|$, $|T_L|$, $T/2$ are plotted with system size $N$ for
two different upper and lower arm ratios near the band edge with $E = (2 + \Delta \xi)$.
In (a), the arm-length ratio is 1, whereas in (b) it is 1/2. The ring site where the drain is attached,
denoted as $N_D$, is marked at the top of (a) and (b).
The red, blue, and green colors represent $\Delta \xi = 0$, $0.0001$, and $0.002$, respectively.
The dashed line in (a) indicates the system size up to which the transmissions increase with $N$
and beyond which they decrease for $\Delta \xi = 0$. For this case, we plot the transmissions
around the band edge for various system sizes in (c), where $E = 2$\,eV is indicated by the dashed line.
In (c), the red, blue, green, magenta, and cyan colors denote
$N = 600$, $700$, $800$, $900$, and $1000$, respectively.
The parameters used for these calculations are $t_0 = 1.2\,$eV and
$t_C = 0.1\,$eV.}
\label{fig3}
\end{figure}
carry a current opposite to the net transport direction, as required by current
conservation. This behavior is well documented in earlier studies of circular current
\cite{Nitzan,zero2}, where the bond-resolved transmission coefficient
$T_{i\rightarrow i+1}$ naturally changes sign. Consequently, the net circular transmission
coefficient $T_C$, defined as the average of the bond transmission coefficients, can also
be positive or negative, even though the total junction transmission probability
$T$ remains strictly positive. Thus, the sign reversal of
the bond and circular transmission coefficients simply reflects the presence of
circulating current components and does not imply any unphysical negative resistance.

\subsection{Scaling relation of the transmission at the band-edge}

For an OQR, electrons propagate within the energy window around $-2t$ to $+2t$. We find
that, apart from the eigenenergies of any QR, the transmissions
are independent of the system size $N$. This is known as ballistic transport, which is the transport regime for a
linear conductor. Anomalous transport is observed around the eigenenergies
of a QR. Let us first concentrate on the band edges, i.e., at $E = \pm 2\,$eV for our chosen
parameter values.

It is worth noting that, for the symmetric configuration, the net circular transmission coefficient is zero.
\begin{figure}
{\centering\resizebox*{7cm}{4.5cm}{\includegraphics{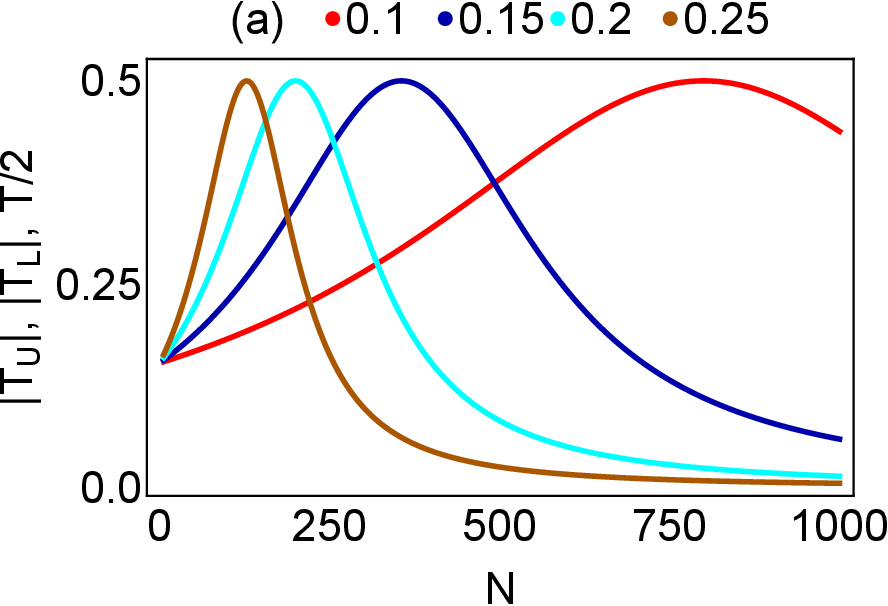}}\\[1em]
\resizebox*{7cm}{4.5cm}{\includegraphics{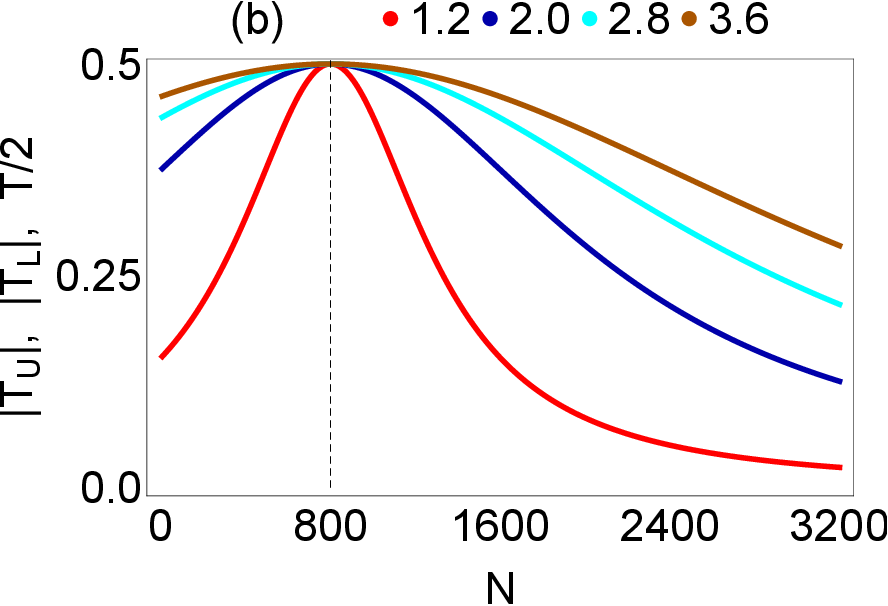}}\par}
\caption{(Color online). System size scaling of the transmissions at the band edge $E = 2\,$eV for a symmetric ring for (a)
different strengths of the ring-to-electrode coupling ($t_C$) with $t_0 = 1.2\,$eV, and (b) different choices of $t_0$ with
$t_C$ fixed at $0.1\,$eV. The choices for $t_C$ in (a) and $t_0$ in (b) are indicated at the top of the respective figures.}
\label{fig4}
\end{figure}
Moreover, for the asymmetric configuration, it has a vanishingly small magnitude near the band edge,
as shown in Figs.~\ref{fig2}(b) and (e). Therefore, we do not focus on the net circular transmission coefficients at
the band edge.

The transmissions for two distinct ring-to-electrode arrangements are plotted in Fig.~\ref{fig3}.
Figure~\ref{fig3}(a) shows the results for a symmetric ring with equal upper and lower arm lengths,
while Fig.~\ref{fig3}(b) presents the results for a ring with an upper-to-lower arm length ratio of $1/2$.
We calculate the transmissions at $E = (2 + \Delta \xi)\,$eV, where $\Delta \xi$ corresponds to a small energy window.
The red, blue, and green curves in Figs.~\ref{fig3}(a) and (b) correspond to $\Delta \xi = 0$, $0.0001$,
and $0.002$, respectively. It is important to note that, in a
symmetric ring, the magnitudes of $T_U$ and $T_L$ are the same, and $T$ is two times larger than each of them. In contrast,
the asymmetric ring exhibits distinct magnitudes of the transmissions
(Fig.~\ref{fig3}(b)), but nearly identical qualitative relationships with $N$.
For this reason, we focus only on one of the transmissions in this case.
These results demonstrate how the transmissions first increase with $N$ at the band edge and subsequently fall
at the band edge and very close to the band edge, as seen from the red and blue curves.
For larger $\Delta \xi$ ($\Delta \xi = 0.002$), the transmissions decrease with $N$,
as shown by the green curve. Thus, the scaling relations at the band edge can be expressed as
\begin{eqnarray}
T_X ~~\mbox{or} ~~T & \sim & N^{\alpha} ~~~\forall~~ N < N^C; \nonumber \\
T_X ~~\mbox{or} ~~T & \sim & N^{-\beta} ~~~\forall~~ N > N^C,
\label{eq5}
\end{eqnarray}
\noindent
where $X = U,~ L$. $\alpha$ and $\beta$ are positive numbers, and $N^C$ is the ring size
\begin{figure}
{\centering\resizebox*{7cm}{4.5cm}{\includegraphics{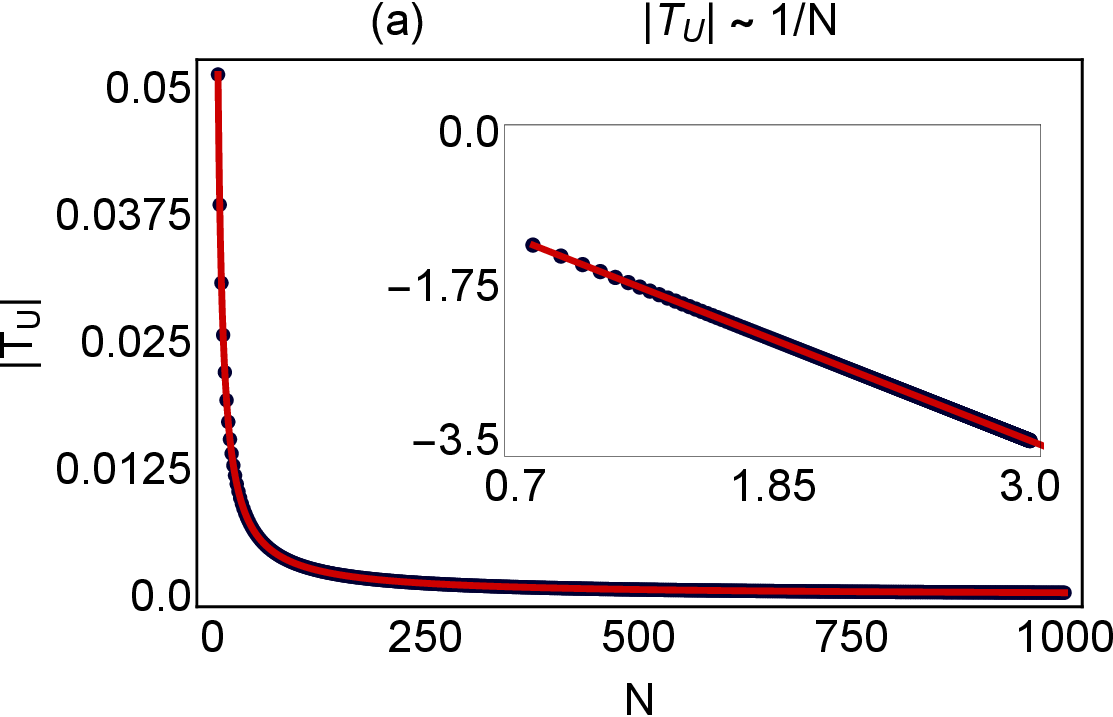}}
\resizebox*{7cm}{4.5cm}{\includegraphics{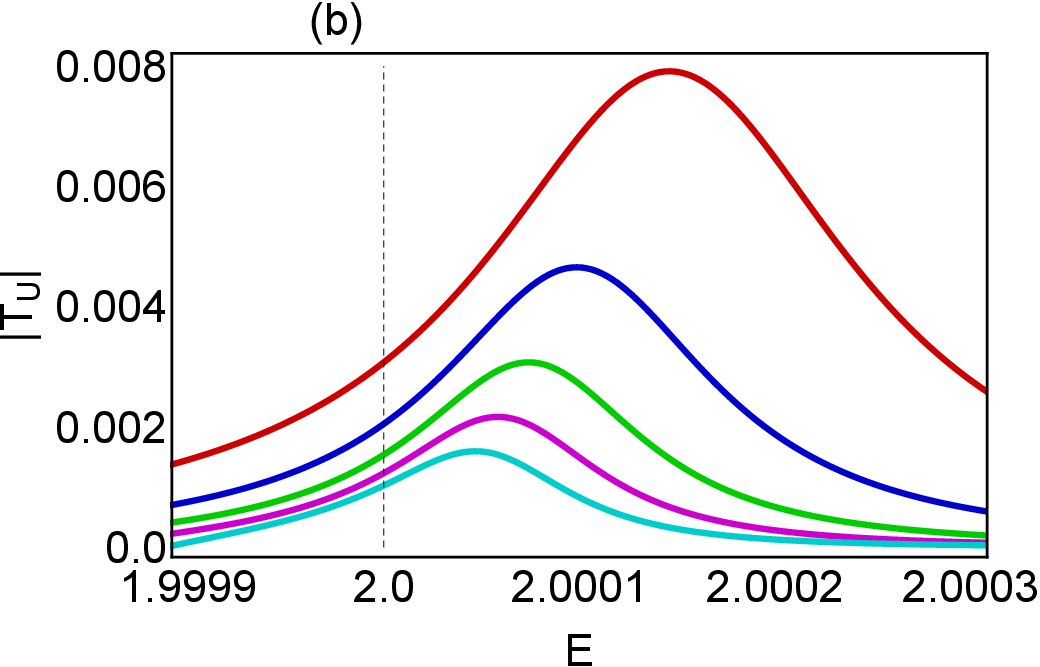}}\par}
\caption{(Color online). (a) Scaling of $|T_U|$ with ring size $N$ at the band-edge $E = 2\,$eV for the most asymmetric ring-to-electrode
configurations. The black dots represent the numerically calculated values. Using these results, we find the scaling relation between
$T_U$ and $N$, which is shown by the continuous red curve. In the inset, the black dots depict the logarithmic plot of the same data.
The red line in the inset represents the fit as $\log |T_U| = -\log N + \text{constant}$. (b) Transmissions around the band edge for different
system sizes. The meaning of the different colors is the same as in Fig.~\ref{fig3}(c). The parameters are the same as in Fig.~\ref{fig3}.}
\label{figX}
\end{figure}
up to which the transmissions increase with $N$. \textbf{\it We classify the growth of
transmissions with $N$ as beyond-ballistic transport.}

In contrast, near the band edge ($\Delta \xi = 0.002$), the transmission scales with $N$ as
\begin{eqnarray}
T_X ~~\mbox{or} ~~T & \sim & N^{-\gamma} ~~~\forall~~ N,
\label{eq5a}
\end{eqnarray}
\noindent
where $X = U,~ L$. Here, $\gamma$ is positive. $\gamma = 1$ indicates diffusive transport;  
$\gamma > 1$ corresponds to sub-diffusive transport; and the regime $0 < \gamma < 1$ is termed
super-diffusive.

To explain the anomalous beyond-ballistic transport, we indicate $N^C$ with a dotted
line for $\Delta \xi = 0$ in Fig.~\ref{fig3}(a). For this case, $N^C = 800$. We now calculate the transmissions around
$E = 2\,$eV for different system sizes, viz.\ $N = 600$, $700$, $800$, $900$,
and $1000$, as shown in Fig.~\ref{fig3}(c). The vertical dotted line in this figure indicates $E = 2\,$eV.
The magnitudes of the transmission maxima for systems with $N\leq N^C$
are comparable, whereas they decrease for larger $N$. As the height of the peaks decreases with $N$,
the maxima positions shift from the right to the left on the energy axis. For these reasons, at the band edge, the transmissions
initially increase with $N$ up to $N^C$ and subsequently decrease. At $\Delta \xi = 0.002$,
the peak maxima drop with $N$, and therefore the transmissions decrease with $N$, as shown by the green curves in
Figs.~\ref{fig3}(a) and (b).

Figure~\ref{fig4} shows how the system parameters affect $N^C$. For a symmetric ring, we plot
the transmissions versus $N$ for various strengths of the ring-to-electrode coupling with fixed $t_0$ in
Fig.~\ref{fig4}(a), and for different choices of $t_0$ with fixed ring-to-electrode coupling strength $t_C$ in
Fig.~\ref{fig4}(b). Figure~\ref{fig4}(a) shows that $N^C$ becomes larger for weaker $t_C$, whereas it does not depend
on $t_0$, as can be seen from Fig.~\ref{fig4}(b). Instead, Fig.~\ref{fig4}(b) indicates that for fixed $t_C$, the
exponents $\alpha$ and $\beta$ depend on $t_0$. When $t_0$ is comparable to the ring hopping ($t$), $\alpha$ and
$\beta$ become larger than 1, resulting in beyond-ballistic transport for $N < N^C$ and diffusive transport for
$N > N^C$. In contrast, for $t_0 \gg t$, the transport behavior tends to be predominantly ballistic.

In Fig.~\ref{figX}(a), we plot the system-size scaling of the transmission for the most asymmetric rings with
$N_D = N$, such that the length difference between the upper and lower arms is maximum.
For the same parameter values as chosen in Fig.~\ref{fig3}, we find that the transmission
$T_U$ decays with $N$ at the band edge $E = 2\,$eV. The nature of the transport is
diffusive, as the scaling relation is
\begin{equation}
T_U \sim N^{-1} ~~~\forall ~~N.
\label{eqscale}
\end{equation}

\noindent
The inset of Fig.~\ref{figX}(a) displays a logarithmic plot, which clearly illustrates the
$N^{-1}$ scaling. This behavior is robust against variations in the ring-to-electrode coupling
strength. To further illustrate this scaling, we plot the variation of $T_U$ around
$E = 2\,$eV for different system sizes in Fig.~\ref{figX}(b). In this plot, we observe that the
resonant peaks shift toward $E = 2\,$eV (indicated by the dashed line) from the right, and the
peak heights decrease as $N$ increases, leading to a power-law decay in transmission with
system size. Figure~\ref{figX} presents the results for $T_U$, but similar patterns are
observed for $T_L$, $T_C$, and $T$.

\subsection{Scaling relation of transmissions within the band}

As already mentioned, apart from the energies corresponding to the eigenvalues
associated with a given $N$, the nature of transport is ballistic.
For a symmetric ring, the transport remains ballistic even at the eigenenergies
within the band. However, we observe beyond-ballistic transport as described
\begin{figure*}
{\centering\resizebox*{17.5cm}{9.5cm}{\includegraphics{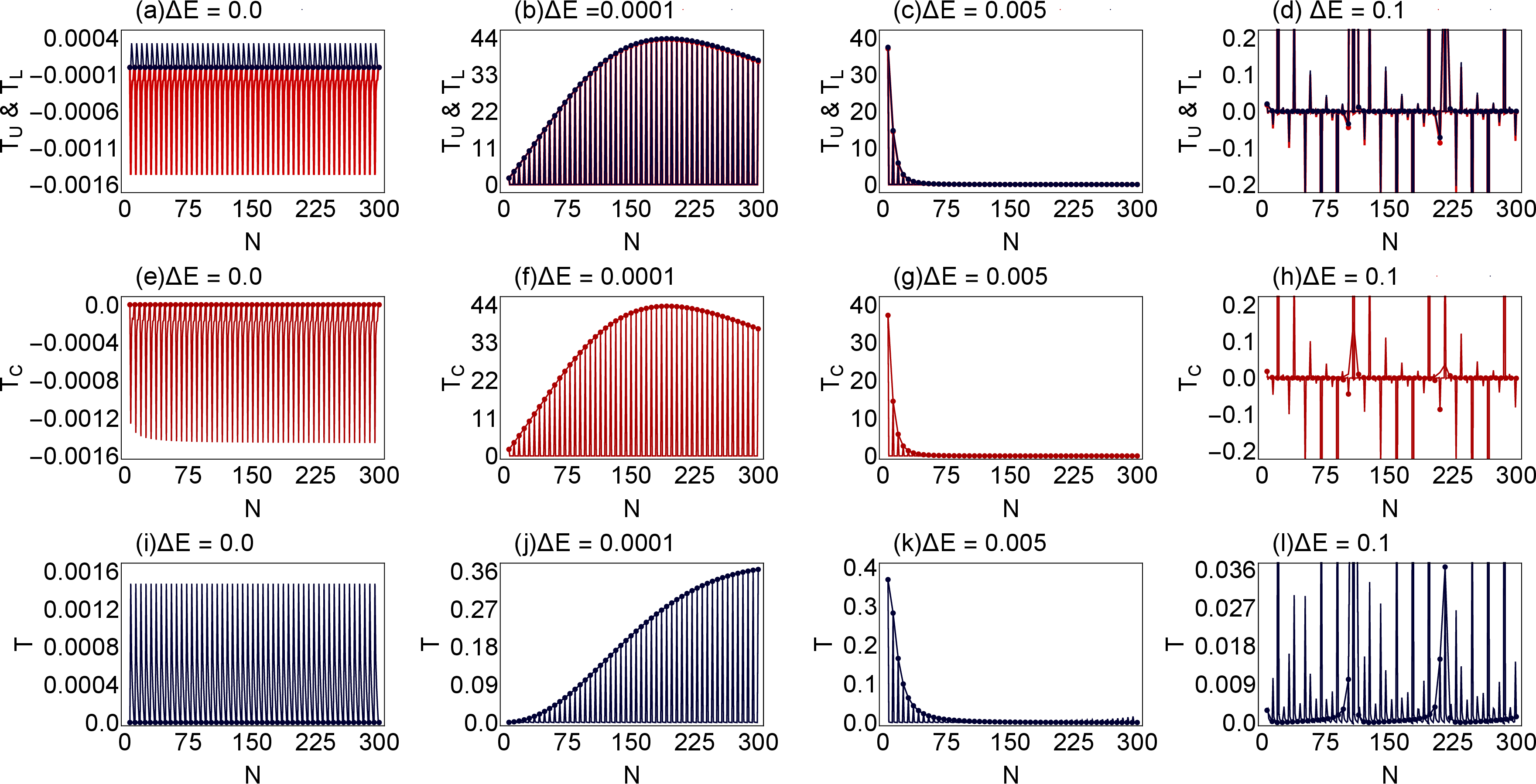}}\par}
\caption{(Color online). The $N$ dependence of the transmission at $E = (1 + \Delta E)\,$eV
with different $\Delta E$'s for the most asymmetric rings. The red and blue colors in the first row represent
the transmissions for the upper and lower arms, respectively. The dots represent the transmission
of the OQRs with $N \in 6\mathbb{N}$. The points are connected to make it evident that
the observed beyond-ballistic scaling arises solely from those rings with $N \in 6\mathbb{N}$, or equivalently, from
systems sharing the common eigenvalue $1$. All the calculations are done for $t_C = 0.2\,$eV and $t_0 = 2.0\,$eV.}
\label{fig5}
\end{figure*}
in Eq.~\ref{eq5}, where the exponents $\alpha$ and $\beta$ are greater than but close
to $1$, provided we consider only those rings that possess the corresponding eigenenergy.
In an asymmetric ring with a commensurate upper-to-lower arm length ratio, the
transmission resonance associated with a doubly degenerate eigenenergy may remain
unsplit or split into two resonant peaks, depending on the ring-to-electrode
configuration. Consequently, in this scenario, transport at the corresponding
eigenenergy remains ballistic.

Now we focus on the most asymmetric junctions. In Fig.~\ref{fig5} we plot the transmissions as a function
of $N$, at and around an energy corresponding to a degenerate eigenvalue. For illustration, we choose
the degenerate eigenenergy $E=1\,$eV, which is common to the $N=6$ ring and all rings whose system
sizes are integer multiples of $6$. The conclusions drawn below are, however, not restricted to
the representative example considered here, but remain valid for any doubly degenerate eigenenergy shared
by a family of quantum rings whose system sizes satisfy a specific integer-multiple relation.
The transmission coefficients for the upper and lower arms are presented in the first row, while the net circular
transmission coefficient is shown in the second row. The third row presents the transmission probability
associated with the junction (transport) current. The results are evaluated at $E = (1 + \Delta E)\,$eV,
with $\Delta E = 0$ for Figs.~\ref{fig5}(a), (e) and (i), $0.0001$ for Figs.~\ref{fig5}(b), (f) and (j),
$0.005$ for Figs.~\ref{fig5}(c), (g), and (k), and $0.1$ for Figs.~\ref{fig5}(d), (h), and (l).
To clarify the role of OQRs with $N \in 6\mathbb{N}$, the corresponding transmission values are
highlighted by dots and connected to form a smooth curve, making their characteristic scaling behavior evident.

As discussed earlier, for $N = 6$, the quantum ring possesses a degenerate eigenenergy at $E = 1\,$eV. Thus,
all rings with $N \in 6\mathbb{N}$ share common degenerate energy levels at $E = \pm 1\,$eV. We find that
the qualitative features of transport are identical for the bond, circular, and junction
transmission quantities. In particular, around the degenerate energies of the bare ring, the bond
and circular transmission coefficients exhibit nearly identical qualitative and quantitative behavior. At $E = 1\,$eV,
the transport is ballistic, since the transmissions are independent of $N$, as seen in Figs.~\ref{fig5}(a), (e), and (i).
For rings with $N \in 6\mathbb{N}$, the transmissions become vanishingly small due to the anti-resonance at
$E = 1\,$eV; hence, the peaks in Figs.~\ref{fig5}(a) and (e) originate from rings with
$N \notin 6\mathbb{N}$. For these rings, $1\,$eV is not an eigenenergy, and thus the transport becomes ballistic.

In Figs.~\ref{fig5}(b), (f), and (j), we examine the transmission at $E = (1 + \Delta E)\,$eV with
$\Delta E = 0.0001\,$eV. The energy offset $\Delta E=10^{-4}\,$eV corresponds to
an energy scale of $0.1\,$meV (or an equivalent bias of approximately $0.1\,$mV), which is readily accessible in mesoscopic
transport experiments~\cite{meV1,meV2,meV3}. Here we observe that up to a certain system size (denoted $N_U^C$ for $T_U$,
$N_L^C$ for $T_L$, $N_C^C$ for $T_C$, and $N_T^C$ for $T$), the transmission increases with $N$, and decreases thereafter.
\begin{figure}
{\centering\resizebox*{8.5cm}{4.5cm}{\includegraphics{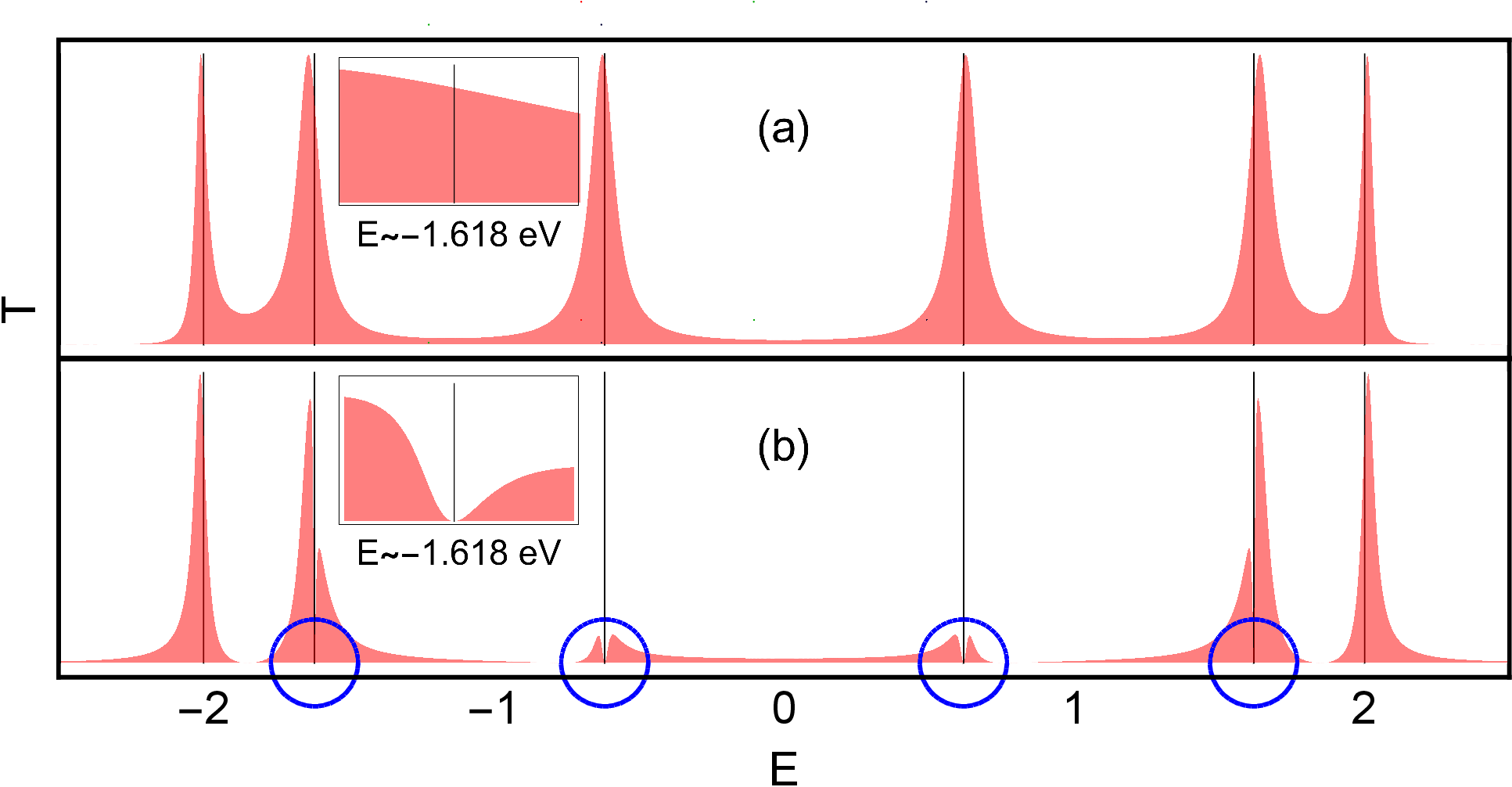}}\par}
\caption{(Color online). Transmission probability as a function of energy for an
ten-site ring with (a) symmetric and (b) most asymmetric ring-to-electrode
configurations. The vertical lines mark the eigenenergies of the bare ring.
The insets show enlarged views of the transmission spectra around the
eigenenergy $-1.1803\,$eV. For the symmetric configuration, no anti-resonance
is observed, whereas for the most asymmetric configuration, transmission
minima with $T\approx 0$ appear at the anti-resonances. The four anti-resonance
points in (b) are highlighted by the blue circles. The calculations are
performed for $t_C=0.5\,$eV and $t_0=2\,$eV.}
\label{fig5x}
\end{figure}
The peaks at small $N$ originate from QRs with $N \in 6\mathbb{N}$, as indicated by the dot-connected curves.
The resulting beyond-ballistic scaling follows a power-series dependence on $N$ of the form
\begin{equation}
T_X ~~\mbox{or}~~T= a_{X_0} + a_{X_1} N^{1} + a_{X_2} N^{2} + a_{X_3} N^{3} + \dots,
\label{SS}
\end{equation}
\noindent
where $X = U,\,L,\,C$. $a_{X_i}$'s ($i = 0,~1,~2,~3,~\dots$) are coefficients.
$a_{X_i}$'s with $i > 2$ become smaller and smaller with $a_{X_{i+1}}/a_{X_i} \sim 10^{-2}$.
For $N > N^C$, the systems have diffusive transport. As our chosen energy is around the
degenerate energy levels for $N \in 6\mathbb{N}$, the bond transmissions for both
the arms have the same sign and almost equal magnitude as we already discussed in Fig.~\ref{fig2}(b).
The transmissions for other $N$ values become very
small compared to the transmission for OQRs with $N \in 6\mathbb{N}$ at $1\,$eV. For $\Delta E = 0.005\,$eV
(Figs.~\ref{fig5}(c) and (g)), we find diffusive transport. For larger values of $\Delta E$ (i.e., $\Delta E = 0.1\,$eV),
\begin{figure}
{\centering\resizebox*{7cm}{10cm}{\includegraphics{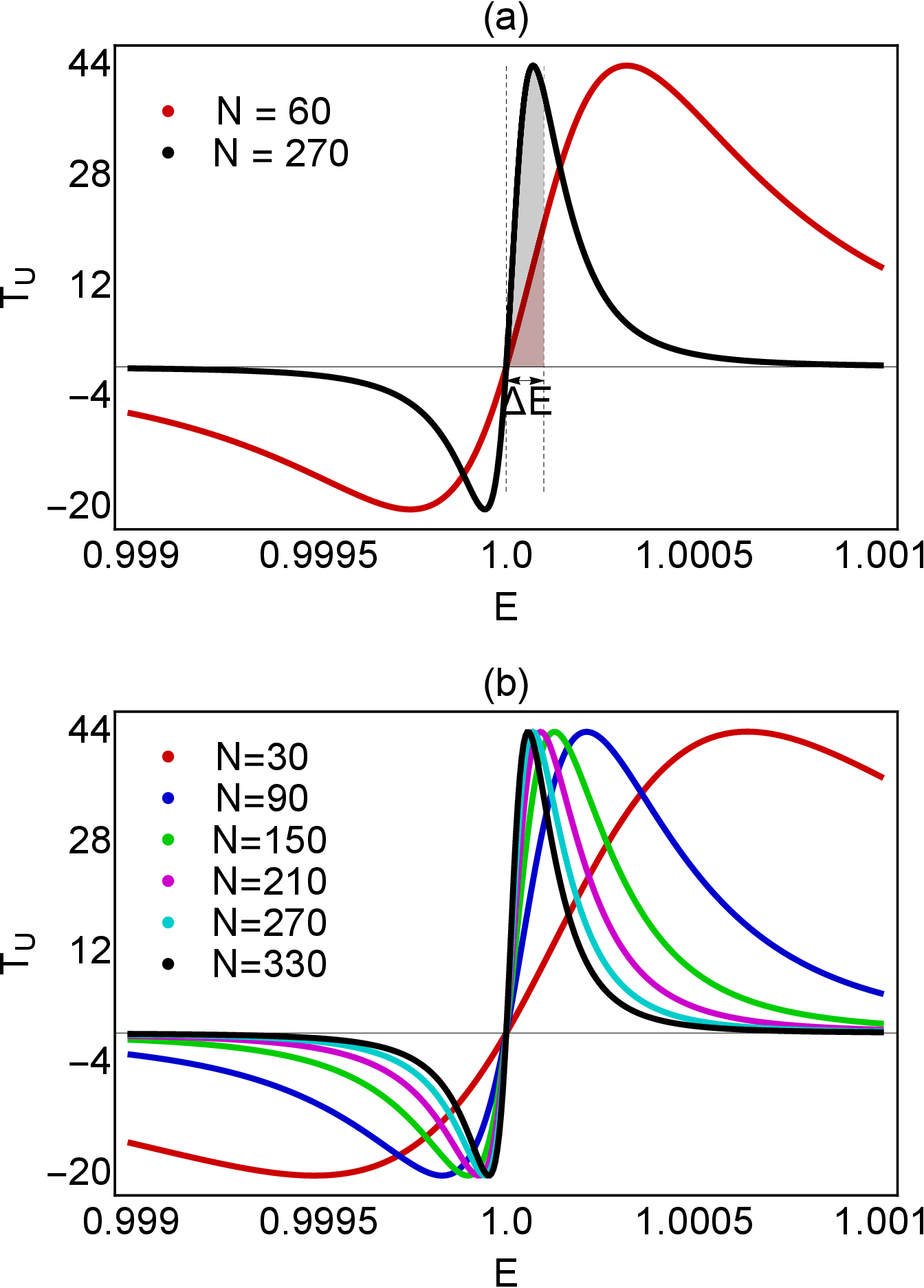}}\par}
\caption{(Color online). (a) $T_U$-$E$ spectra around $E =1\,eV$ for $N = 60$ and $270$ for the same parameter values of
Fig.~\ref{fig5}. The energy window $1\leq E \leq 1 + \Delta E$, with $\Delta E = 0.0001$ is shown by the shaded region.
(b) $T_U$-$E$ spectra around $E =1\,eV$ for different OQRs with $N \in 6\mathbb{N}$ for
the same parameter values of Fig.~\ref{fig5}.}
\label{fig6}
\end{figure}
we find ballistic transport as shown in Figs.~\ref{fig5}(d), (h), and (l) where the transmissions are mostly dominated by the
ring sizes $N \notin 6\mathbb{N}$. The results shown in Fig.~\ref{fig5} are valid for
any degenerate eigenenergy of an isolated ring, where an asymmetric ring-to-electrode
configuration gives rise to an anti-resonance at that energy. As an illustration, we consider
a ring with $N=10$, and plot the transmission probability as a function
of energy for the symmetric and most asymmetric ring-to-electrode configurations in
Fig.~\ref{fig5x}. For $N=10$, the isolated ring possesses four doubly degenerate eigenenergies,
located at $E=\pm1.61803\,$eV and $\pm0.618034\,$eV, which are indicated by the vertical lines.
For the symmetric ring-to-electrode configuration, no anti-resonance is observed at these
degenerate energies, as shown in Fig.~\ref{fig5x}(a) and its inset. In contrast, for the
most asymmetric configuration, pronounced anti-resonances appear precisely at these
degenerate eigenenergies of the isolated ring, as highlighted by the blue circles in
Fig.~\ref{fig5x}(b). Consequently, the anomalous (beyond-ballistic) transport behavior
is observed in the vicinity of these degenerate energies.

There exist further asymmetric configurations with $N_D = N -1,~N-2,~\dots$ that exhibit similar anomalous scaling.

\subsection{To find $\Delta E$}

The choice of $\Delta E$ is very important to find beyond-ballistic transport in an asymmetrically connected ring.
Let us first understand the role of $\Delta E$ to have the anomalous system-size scaling of $T_U$.
\begin{figure}
{\centering\resizebox*{8cm}{3cm}{\includegraphics{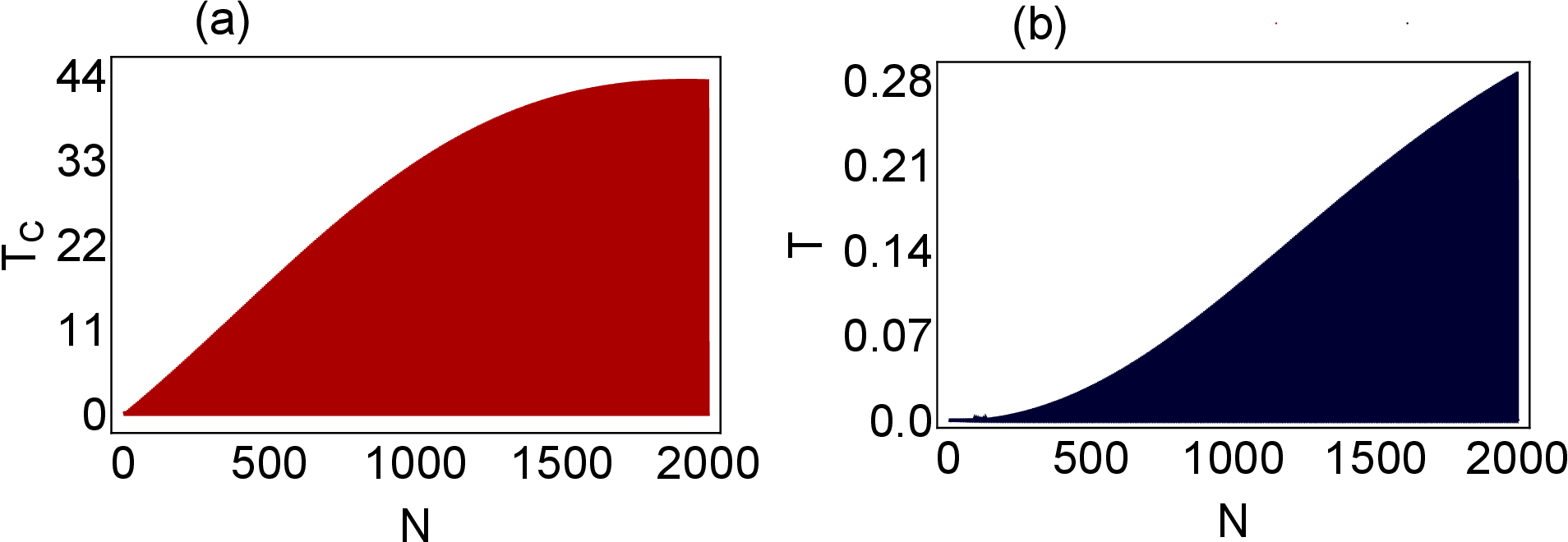}}\par}
\caption{(Color online). System-size dependence of (a) the net circular transmission coefficient
$T_C$ and (b) the junction transmission probability $T$ for the most asymmetric
ring-to-electrode configuration with $\Delta E=0.00001\,$eV. All other parameters are
identical to those used in Fig.~\ref{fig5}. Reducing $\Delta E$ from $0.0001\,$eV to
$0.00001\,$eV extends the beyond-ballistic regime to significantly larger system sizes,
which persists around $N = 2000$, the largest system size considered.}
\label{fig6new}
\end{figure}
We select two values of $N$ from Fig.~\ref{fig5}(b), such that for the first one $T_U$ is increasing with $N$ and for the second
one it is decreasing with $N$. For example, we choose $N = 60$ and 270,
respectively, as these ring sizes are integer multiples of 6. We plot the $T_U$
around $E = 1\,$eV for $N = 60$ and $270$ in Fig.~\ref{fig6}(a). The OQRs with $N \in 6\mathbb{N}$ have degeneracy
around $1\,$eV that is split due to asymmetric ring-to-electrode connection. We only concentrate on the
right side of $E = 1\,$eV as we choose $E = 1 + \Delta E$ ($\Delta E$ is positive)
for our calculations of Fig.~\ref{fig5}. Let us define a quantity $\Delta E^N$, such that at
$1 + \Delta E^N$ we have the maximum of $T_U$ on the
right side of $E = 1\,$eV for a particular $N$. In other words, $\Delta E^N$ is the splitting of a
degenerate energy level due to asymmetric ring-to-electrode configuration
from the degenerate energy of an isolated ring. From Fig.~\ref{fig6}(a), we find that
$1 + \Delta E < 1 + \Delta E^{60} ~~~~\mbox{but}~~~~ 1 + \Delta E > 1 + \Delta E^{270}$.
As around $E = 1\,$eV, all resonant peaks of $T_U$ corresponding to different $N$ with $N \in 6\mathbb{N}$
have almost the same magnitude (as shown in Fig.~\ref{fig6}(b)). Therefore, we can conclude that, as long as our
chosen $\Delta E$ is less than $\Delta E^N$, $T_U$ increases with $N$ (as we see for $N = 60$). But when
$\Delta E$ becomes greater than $\Delta E^N$, $T_U$ decreases with $N$ (as for $N = 270$).
For $N = N_U^C$, $\Delta E$ becomes equal to $\Delta E^N$.
In a similar way, we can explain the transport behavior of $T_L$, $T_C$, and $T(E)$ up to $N = N_L^C$,
$N = N_C^C$, and $N = N_T^C$, respectively. It is worth noting that, because the splittings of
the resonant peaks at the left and right sides of the anti-resonance are not the same, $\Delta E^N$ will be different for both sides.

The above analysis also explains how the extent of the beyond-ballistic regime can be controlled.
If a smaller value of $\Delta E$ is chosen, the condition $\Delta E<\Delta E^N$ remains valid up to
larger system sizes. Consequently, the beyond-ballistic scaling persists over a much wider range
before the crossover to the diffusive regime occurs. To demonstrate this, in
Fig.~\ref{fig6new} we plot the net circular transmission coefficient $T_C$ and the junction
transmission probability $T$ as functions of the system size $N$, keeping all the parameters
the same as those used in Fig.~\ref{fig5}, except that $\Delta E$ is reduced from $0.0001\,$eV to
$0.00001\,$eV~\cite{meV1,meV2,meV3}. We find that both $T_C$ and $T$ continue to exhibit
beyond-ballistic scaling up to approximately $N = 2000$, the largest system size considered,
and are expected to persist for even larger system sizes. These results clearly demonstrate that the
beyond-ballistic regime is not restricted to a narrow range of system sizes; rather, its extent
can be systematically controlled by the choice of $\Delta E$ relative to the resonance splitting
$\Delta E^N$.

\subsection{Role of $N$ and $t_C$ on $\Delta E^N$}

\begin{figure}
{\centering\resizebox*{7cm}{9cm}{\includegraphics{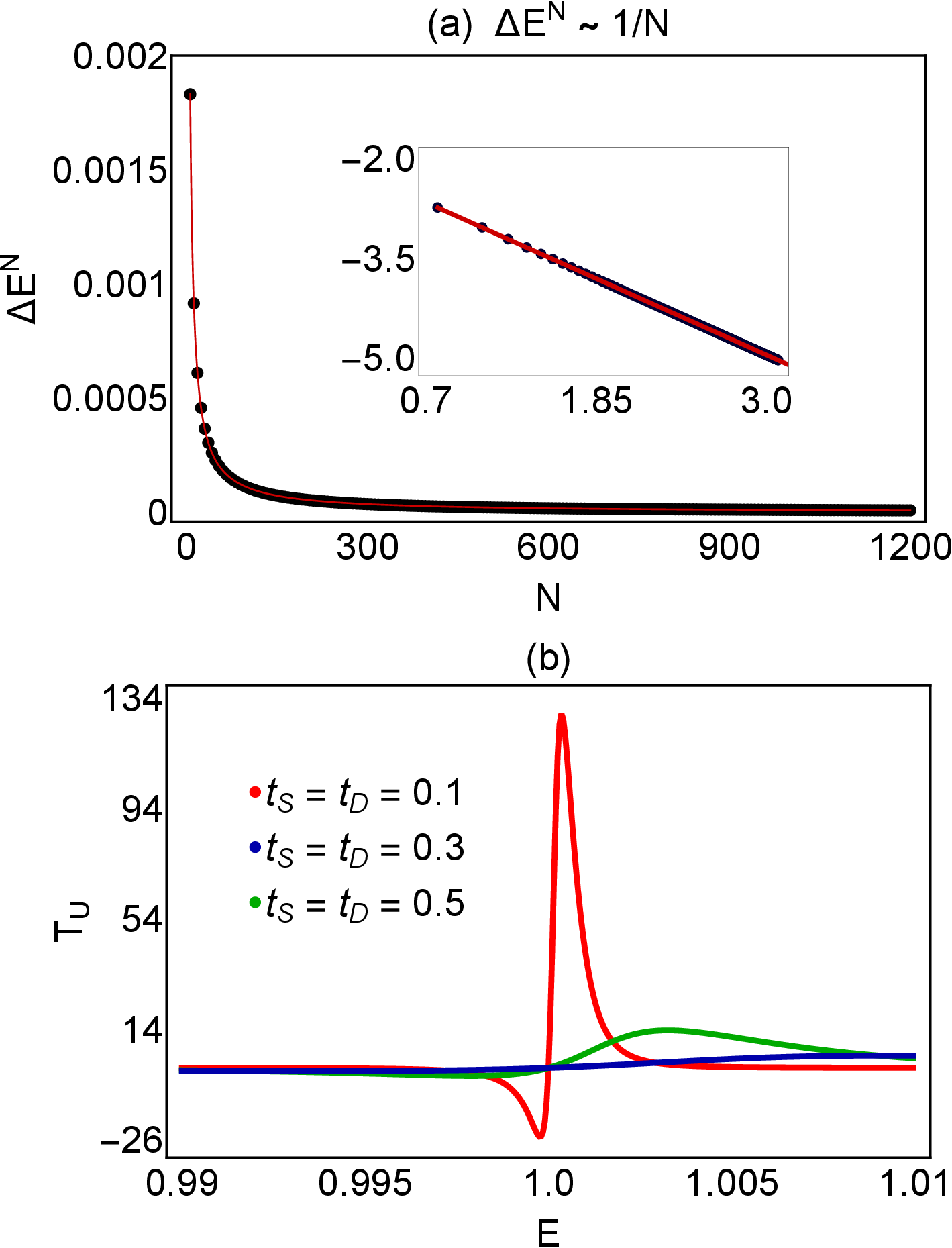}}\par}
\caption{(Color online). (a) Variation of $\Delta E^N$ with ring size $N$ around $E = 1\,$eV. The meanings of the black dots and red curve are
the same as mentioned in Fig.~\ref{fig5}. In the inset, the black dots illustrate the logarithmic plot of the same data. The red line in
the inset shows the fit as $\log \Delta E^N = - \log N +$ constant. Here we choose $t_C = 0.1\,$eV and $t_0 = 1\,$eV. (b) Resonant peaks
of $T_U$ around the degenerate energy $E_n = 1\,$eV and $N = 30$ for different ring-to-electrode coupling strengths.}
\label{fig7}
\end{figure}

From the results discussed so far, we find that to have beyond-ballistic scaling behavior, we need to set
the Fermi energy at $E_m \pm \Delta E$, where $E_m$ is the degenerate energy
corresponding to the isolated ring and $\Delta E$ is less than $\Delta E^N$.
The splitting of the transmission resonance associated with a doubly degenerate
eigenenergy due to the asymmetric ring-to-electrode configuration is very small
(of the order of a few meV or less) and decreases with system size $N$ as $1/N$,
as shown in Fig.~\ref{fig7}(a) and its inset.
Now we concentrate on making
$\Delta E^N$ reasonable enough for experimental observation. One of the possible ways
to obtain larger splitting is by controlling the ring-to-electrode coupling.
In Fig.~\ref{fig7}(b) we plot the $T_U$ around a degenerate energy (say, $1\,$eV for
$N = 30$) for different values of ring-to-electrode coupling strengths.
As we can see here, $\Delta E^N$ increases with $t_C$. Therefore, we can choose
a higher $\Delta E$ for larger $t_C$. But with increasing $t_C$, the peak heights of $T_U$
decrease. As a result, for a fixed $\Delta E$ and larger $t_C$,
the corresponding beyond-ballistic behavior becomes a small-scale phenomenon ($N^C_X$s, $X = U,~L,~C,~T$, become
smaller). So, we can choose a moderate $t_C$ to have the beyond-ballistic scaling relation for a reasonable
$\Delta E$. {\it From our study we find that, if the transmission resonance associated with a doubly degenerate
eigenenergy splits because of the asymmetric ring-to-electrode configuration, the peak heights and $\Delta E^N$ are
independent of the positions of the electrodes. Hence, the estimated $\Delta E^N$
does not depend on the specific ring-to-electrode connection. This also indicates the robustness of our study.}

In a similar way, we can fix $t_C$ for $T_L$, $T_C$, and $T$.

\subsection{Effects of disorder, temperature, dephasing, and subbands}

To gain deeper insight into the robustness of anomalous transport in quantum rings,
it is essential to examine the role of disorder. In realistic systems, on-site energy
fluctuations arising from impurities or structural imperfections are unavoidable and
can significantly influence the nature of transport.
Here, we examine the Aubry-Andr\'{e} model~\cite{dis1,dis2,dis3,dis4,dis5} of correlated disorder where
\begin{figure}
{\centering\resizebox*{4cm}{2.8cm}{\includegraphics{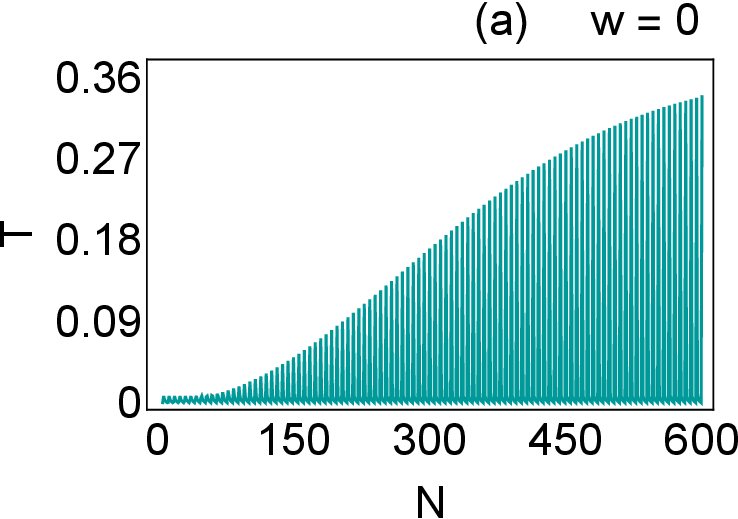}}
\resizebox*{4cm}{2.8cm}{\includegraphics{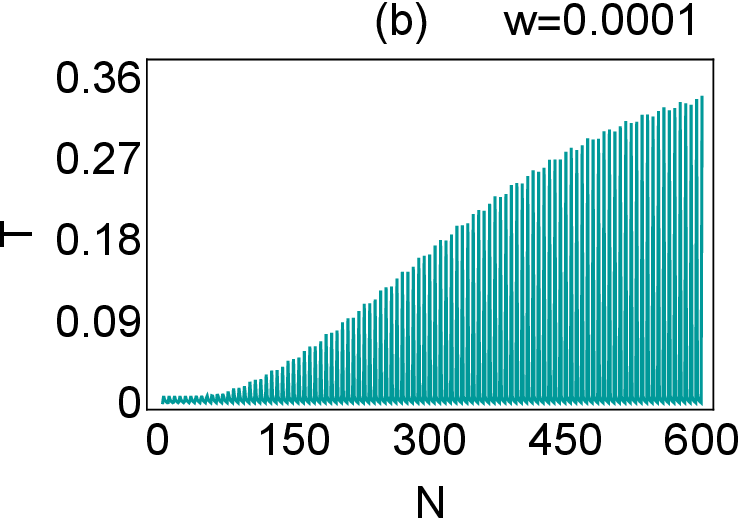}}
\resizebox*{4cm}{2.8cm}{\includegraphics{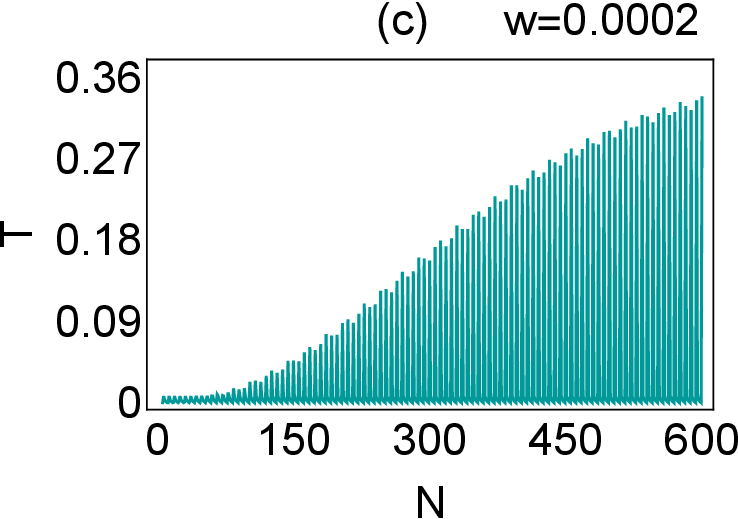}}
\resizebox*{4cm}{2.8cm}{\includegraphics{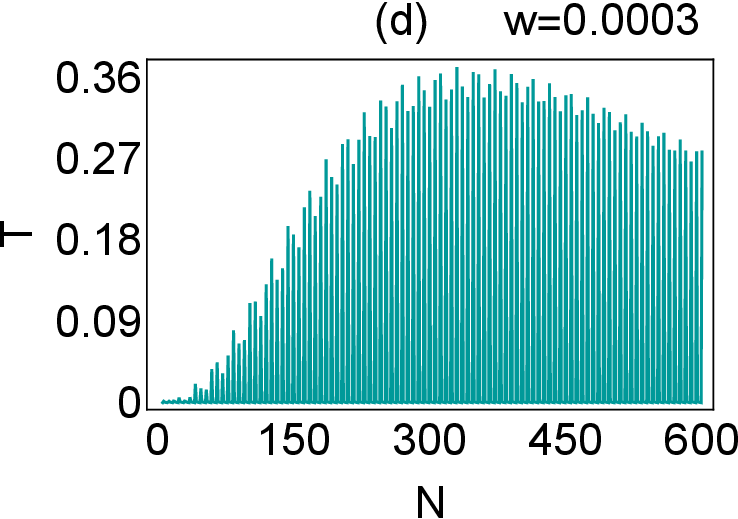}}\par}
\caption{(Color online). Variation of $T$ with ring size $N$ in the presence of AA modulation of
the on-site potential with different strengths of disorder. Here we consider the most asymmetric ring-to-electrode
configuration with $t_C = 0.2\,$eV and $t_0 = 2\,$eV,
and choose the energy near a doubly degenerate eigenenergy,
$E = (1 + 0.0002)\,$eV.}
\label{fig7a}
\end{figure}
the impurities in the ring are included by choosing the on-site energies of the ring in the form of
$\epsilon_i = w \cos(2\pi b i)$, where
$w$ is the impurity strength and $b$ is an irrational number. We set $b = (1 + \sqrt{5})/2$.
The ordered ring is represented by $w = 0$.

To investigate the effect of disorder, one can also look at random site energies rather than
``correlated" disorder. However, in that scenario, we need to take the average over a huge
number of different disordered configurations. To avoid this, we disregard random
distribution of site energies, and we expect identical physical outcomes in both uncorrelated and correlated
scenarios in our current discussion.
We have illustrated the impact of disorder on the system-size variations of the
transmission probability in Fig.~\ref{fig7a}. We only focus on the most asymmetric ring-to-electrode configuration.
Figure~\ref{fig7a}(a) shows the result for the ordered ring. Here we set the energy at $(1 + \Delta E)\,$eV.
As $1\,$eV is a degenerate energy level for rings with $N \in 6\mathbb{N}$,
we find beyond-ballistic transport (up to $N^C$, say) near degeneracy for the ordered case as shown in Fig.~\ref{fig7a}(a).
Figures~\ref{fig7a}(b), (c), and (d) represent disordered rings with $w = 0.0001\,$, $0.0002\,$, and
$0.0003\,$, respectively. We find that, in the presence of small
disorder such that $w < \Delta E$ or $w \sim \Delta E$, the beyond-ballistic behavior remains intact.

\begin{figure}
{\centering\resizebox*{7cm}{4.7cm}{\includegraphics{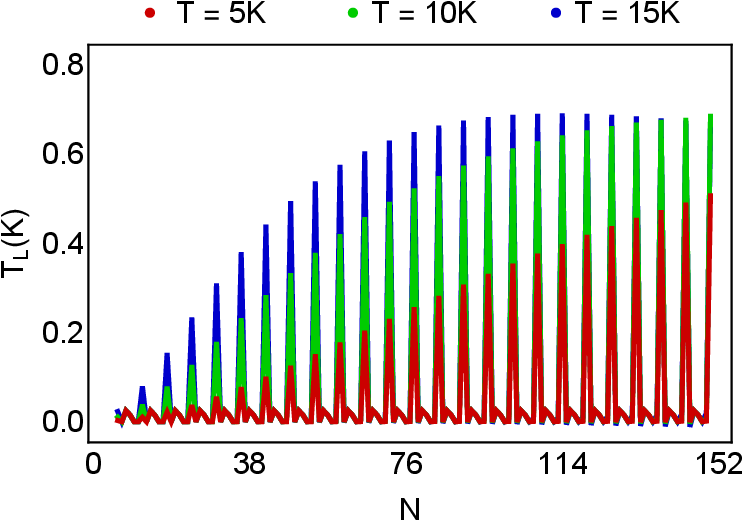}}\par}
\caption{(Color online). Variation of the transmission $T_L$ with the number of sites
$N$ is shown for different finite temperatures, as indicated at the top of the figure. The calculations
are performed for the most asymmetric ring-to-lead configuration with $t_0 = 2\,$eV and $t_C = 0.6\,$eV,
while the energy is fixed at $1\,$eV.}
\label{fig8a}
\end{figure}

So far, all results have been obtained at zero temperature, where transport is governed
solely by quantum coherence. However, in realistic situations, finite temperature leads to thermal
broadening of electronic states and modifies the energy distribution of carriers. Incorporating
temperature effects thus provides a more practical picture.
In the linear response regime, i.e., when the voltage and temperature differences between
the two leads are sufficiently small, the transmission probability can be evaluated
as~\cite{tem1,tem2,tem3}:
\begin{equation}
T(K) = \int_{-\infty}^{\infty} T(E) \left(-\frac{\partial f}{\partial E}\right) \,dE.
\label{eqT2}
\end{equation}
\noindent
The results are displayed in
Fig.~\ref{fig8a} and clearly demonstrate the persistence of anomalous beyond-ballistic scaling at finite
temperatures. At non-zero temperature, the Fermi energy can be chosen at a degenerate
eigenenergy, while thermal broadening accounts for contributions from a
finite energy window. Thermal broadening suppresses the anti-resonance. Consequently, the choice of $\Delta E$ is no longer
relevant in this regime. It is further observed that the characteristic length $N^C$, over
which beyond-ballistic scaling is sustained, decreases systematically with increasing temperature,
as illustrated in Fig.~\ref{fig8a}.

In addition to disorder and temperature, quantum coherence plays a crucial role in
determining transport properties. In realistic systems, interactions with the environment inevitably lead
to decoherence, which suppresses phase-coherent effects such as interference. To investigate the influence
of dephasing on the scaling behavior, we employ the B\"{u}ttiker probe approach, wherein fictitious voltage
probes are introduced to mimic inelastic scattering processes. This framework allows us to systematically
analyze how the gradual loss of phase coherence modifies the transmission characteristics of the quantum ring.
In the B\"{u}ttiker probe method, fictitious voltage probes are attached
to each lattice site of the ring except the sites where the source and
drain are connected. These probes are coupled to the system with a
\begin{figure}
{\centering\resizebox*{7cm}{14cm}{\includegraphics{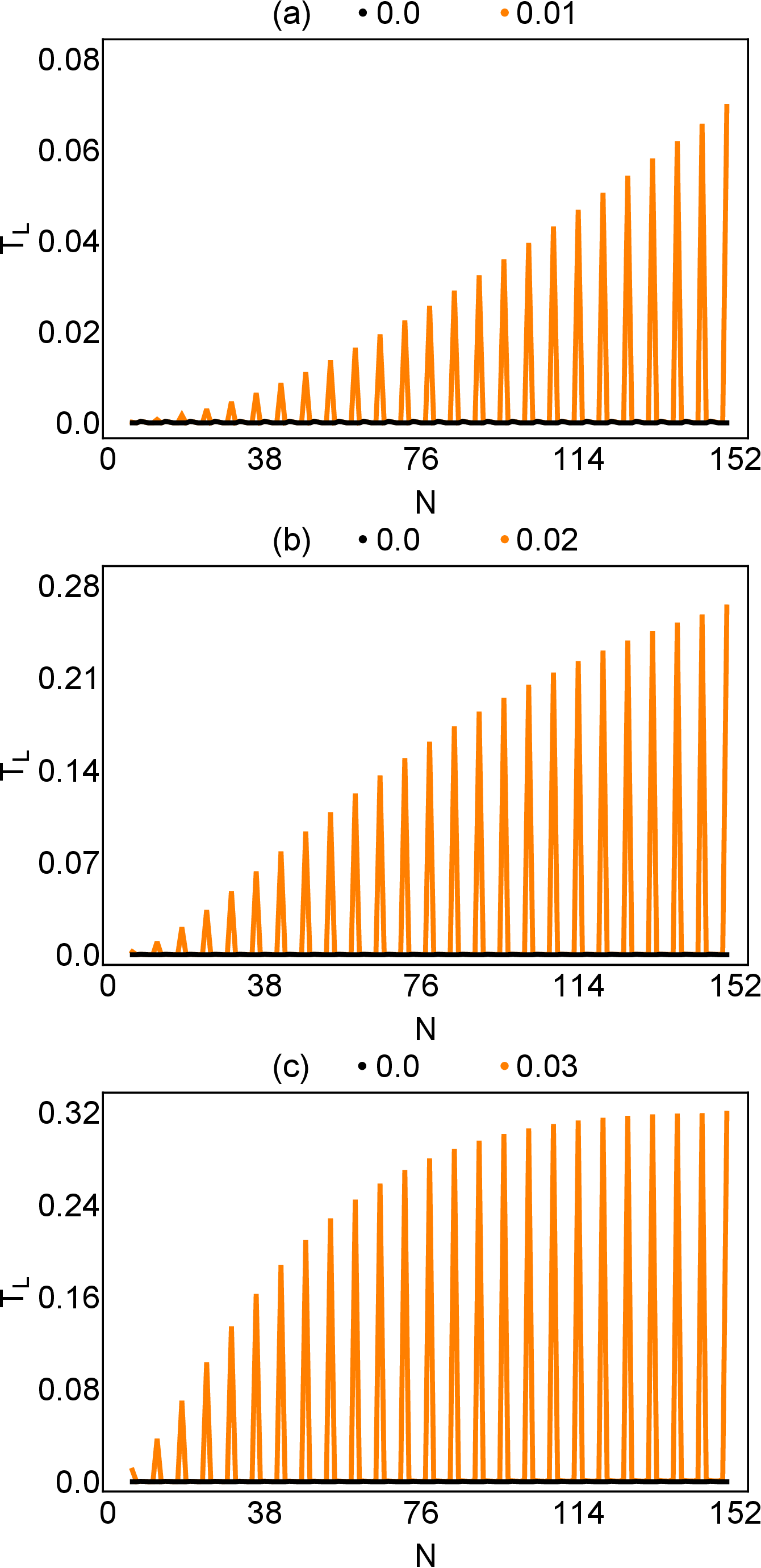}}\par}
\caption{(Color online). Effects of dephasing on the variation of $T_L$ with $N$.
Each panel shows a superposition of the zero-dephasing case (black curve) with the corresponding non-zero
dephasing cases for clarity. The calculations are performed for the most asymmetric ring-to-lead
configuration with $t_0 = 2\,$eV and $t_C = 0.2\,$eV, while the energy is fixed at $E = 1.0\,$eV.}
\label{fig8b}
\end{figure}
uniform strength $t_{deph}$, which serves as the dephasing parameter.
Instead of determining the probe electrochemical potentials
self-consistently from the zero-current condition, we assume a linear
voltage drop across the ring. Specifically, for a finite bias applied
between the source and drain, we set $V_S=V_0$ and $V_D=0$ and assume
that the electrochemical potentials of the virtual probes vary linearly
between these two values. Following the method described in
Ref.~\cite{mpprb}, we calculate the transmission in the presence of
dephasing. Fig.~\ref{fig8b} illustrates the scaling of transmission
with system size in the presence of dephasing. It is evident that
beyond-ballistic transport persists even at finite dephasing strength.
Similar to the case of finite temperature, dephasing destroys the
anti-resonance at degenerate energies, leading once again to the
emergence of anomalous scaling behavior at these points.

Unlike conventional cases, the transmission here is found to increase with external
perturbations. This counter-intuitive behavior has been discussed previously~\cite{mpprb}. As
the detailed exploration of this aspect lies beyond the scope of the present work, it will be
addressed in a future study.

In the present work, we consider a strictly one-dimensional quantum ring where electron motion is confined
along a single path, and hence no subband structure arises. To assess the possible influence of dimensionality,
we have examined a corresponding two-dimensional square-lattice ring, which naturally supports multiple
propagation channels due to the presence of subbands. With zero on-site energy, a hopping strength of $1\,$eV,
and an odd number of atomic sites per side, a common eigenenergy is found near $E = 1\,$eV. We analyzed the
size-dependent behavior of the junction transmission at and around this energy for different drain contact
positions. The results for the two-dimensional case exhibit conventional oscillatory variations in transmission
characteristic of ballistic transport, without any indication of anomalous scaling. This comparison
suggests that the beyond-ballistic behavior reported here is an intrinsic feature
of strictly one-dimensional quantum rings and arises only when anti-resonances appear at the
repeating degenerate eigenenergies of integer-multiple system sizes. Such a condition is not
realized in the two-dimensional ring considered here.

\subsection{Total current}

We now investigate whether the currents exhibit anomalous scaling with system size.
\begin{figure}
{\centering\resizebox*{7cm}{4.7cm}{\includegraphics{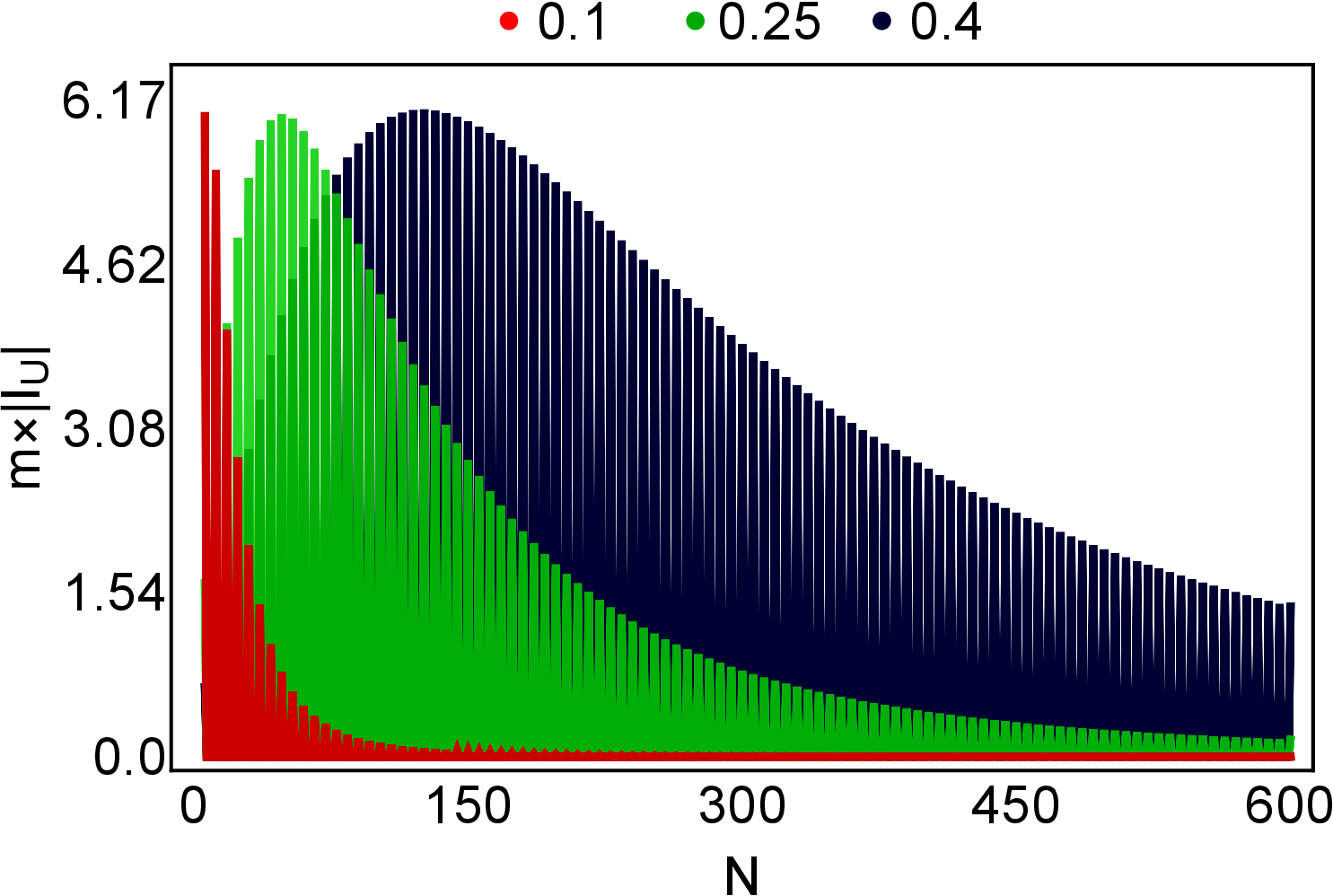}}\par}
\caption{(Color online). Variation of current $|I_U|$ with $N$ for different
ring-to-electrode coupling strengths $t_C$. For clarity, the currents are
scaled by factors $m = 1$, $6$, and $15$ for $t_C=0.1$, $0.25$, and $0.4\,$eV,
respectively. The current is calculated in units of $\mu$A for an applied
bias of $1\,$mV, and the Fermi energy is set at $1\,$eV.}
\label{fig9}
\end{figure}
To calculate the current, we integrate the bond transmission coefficient
over the energy window. As the peak and dip heights are unequal,
we can set the Fermi energy exactly at the degenerate energy level and integrate over a small energy
window to study the scaling relationships. The applied voltage should satisfy $eV < \Delta E^N$.
In Fig.~\ref{fig9}, we plot $|I_U|$ vs $N$ for $t_C = 0.1$, $0.25$, and $0.4\,$eV,
shown by red, green, and blue colors, respectively. $|I_U|$ is multiplied by $1$, $6$, and $15$ for
$t_C = 0.1$, $0.25$, and $0.4\,$eV, respectively, for better presentation.
The currents are calculated for $1\,$mV. We set the Fermi energy corresponding
to the degenerate energy for $N \in 6\mathbb{N}$. Anomalous growth of current
with system size $N$ is also seen here. For higher ring-to-electrode coupling strength,
the beyond-ballistic growth of the current around the anti-resonance persists up to larger $N$,
indicating an extended beyond-ballistic regime.

\section{Connection to experiments}

To experimentally measure the circular current—which is of significant theoretical interest—
several approaches can be considered based on the associated magnetic field it generates.
It has been noted that a circular current can induce a strong local magnetic field. For instance,
Tagami and Tsukada demonstrated that in a T-shaped tape-porphyrin molecular wire, the current
circulating within the molecular loop induces a local magnetic field of about $0.1\,$T at a bias
voltage of $1.2\,$V~\cite{exp1}. By analyzing the spectral response of magnetic ions placed on or near
the ring, the corresponding circular current can be detected—similar to the observation of
magnetic shielding and deshielding in NMR spectra of aromatic molecules~\cite{exp2}. Another possible
method involves measuring the response of the magnetic moment induced in the ring to an external magnetic
field, which can then be used to determine the circular current~\cite{Nitzan1}.

Furthermore, the energy offsets considered in the present work are
experimentally realistic. In particular, mesoscopic transport
experiments have been performed with energy resolutions in the
microelectronvolt ($\mu$eV) range~\cite{meV1,meV2,meV3}. As
the energy offsets $\Delta E$ considered in the present work are
significantly larger than the experimentally demonstrated
microelectronvolt energy scale, they are well within the experimentally
accessible regime. Consequently, the
parameter range explored in the present work should be readily
achievable in mesoscopic quantum-ring devices.

\section{Closing Remarks}

We have studied, for the first time, the system-size scaling of the bond transmission coefficients,
the net circular transmission coefficient, and the overall junction transmission probability in an open quantum ring (OQR).
Unlike a linear chain, a QR possesses doubly degenerate eigenenergies, some of which are common to
families of rings whose system sizes satisfy specific integer-multiple relations. Around these
eigenenergies, the transmission resonances exhibit maxima with the same height, while their widths
decrease with increasing ring size $N$. As a result, anomalous (beyond-ballistic) scaling, characterized
by an increase of the transmission with increasing system size, emerges in an OQR in the vicinity of
these eigenenergies. Below we summarize our key findings. \\

\noindent
$\bullet$ Although the bond transmission coefficients and the net circular transmission coefficient
are quantitatively different, they exhibit nearly identical qualitative scaling behavior with
system size. Likewise, although the transmission coefficients within the ring and the overall
junction transmission probability differ quantitatively, they display similar qualitative
dependence on the ring size $N$. Similar to linear conductors, ballistic transport is observed
throughout the conducting energy window, except in the vicinity of the eigenenergies of the
isolated quantum ring.\\

\noindent
$\bullet$ Beyond-ballistic transport is observed in the vicinity of the non-degenerate
band-edge eigenenergies for the symmetric ring-to-electrode configuration, since quantum
rings with an even number of sites share the same band-edge eigenenergies. At larger
system sizes, the transport crosses over to the diffusive regime. The extent of the
beyond-ballistic regime can be effectively tuned by the ring-to-electrode coupling strength.\\

\noindent
$\bullet$ Within the band, all the eigenenergies of an isolated quantum ring (except the band-edge
states) are doubly degenerate. For an asymmetric ring-to-electrode configuration, the
transmission resonance associated with each doubly degenerate eigenenergy splits into two
resonant peaks separated by an anti-resonance. Beyond-ballistic transport is observed in the
vicinity of these doubly degenerate eigenenergies under asymmetric ring-to-electrode
configurations.\\

\noindent
$\bullet$ We have examined the influence of various physical factors that typically suppress
coherent transport. Remarkably, the beyond-ballistic scaling remains robust under moderate
disorder, finite temperature, and dephasing introduced through the B\"{u}ttiker probe
formalism—although the characteristic scaling length gradually decreases with increasing temperature or
dephasing strength. In contrast, when the system is extended to a two-dimensional lattice
supporting subband formation, the transmission exhibits only conventional ballistic fluctuations.
These results confirm that the observed beyond-ballistic behavior is an intrinsic feature of
coherent one-dimensional quantum-ring transport.\\

Before concluding, we emphasize that the anomalous (beyond-ballistic)
scaling identified in this work originates from the interplay between
the doubly degenerate eigenenergies of the isolated quantum ring,
arising from its periodic boundary condition, and the anti-resonances
generated by asymmetric ring-to-electrode coupling in the corresponding
open system. This mechanism is unique to open quantum rings and has no
counterpart in open quantum junctions with linear conductors. We hope
that the present work will stimulate further theoretical and
experimental studies of quantum transport and system-size scaling in
open quantum rings under more general conditions, including applied
magnetic fields, spin--orbit interactions, disorder, many-body effects,
and stronger dephasing mechanisms.


\section*{ACKNOWLEDGMENT}

MP is thankful to DST-SERB, India (File number: PDF/2022/001168) for providing her research fellowship.

\end{document}